\documentclass[aps,prb,twocolumn,superscriptaddress,floatfix,nofootinbib]{revtex4-2}
\usepackage{graphicx}
\usepackage{amsmath,amssymb,bm}
\usepackage{multirow}
\usepackage{xcolor}
\usepackage[colorlinks=true,linkcolor=blue,citecolor=blue,urlcolor=blue]{hyperref}
\hypersetup{pdftitle={Systematic extinctions in inelastic neutron scattering
from molecular spin clusters}}

\newcommand{\Q}{\mathbf{Q}}
\newcommand{\Rv}{\mathbf{R}}

\begin{document}

\title{Systematic extinctions in inelastic neutron scattering from
molecular spin clusters}

\author{Shadan Ghassemi Tabrizi}
\email{s.ghassemi-tabrizi@hzdr.de}
\affiliation{Computational System Sciences, Technische Universit\"at Dresden, 01187 Dresden, Germany}
\affiliation{Center for Advanced Systems Understanding (CASUS), Am Untermarkt 20, 02826 G\"orlitz, Germany}

\date{\today}

\begin{abstract}
Inelastic neutron scattering on a single crystal resolves how the intensity of a magnetic
transition of a molecular spin cluster varies with the momentum-transfer vector $\Q$. The
point group fixes that dependence completely when a single symmetry species mediates the
transition and the local spin operators contain only one occurrence of that species. Functions for such universal $\Q$ dependences have been
tabulated. As we show here, even when these conditions are not fulfilled, the point group
still fixes, at a given geometry, momentum transfers at which the intensity vanishes for
every Hamiltonian that has the symmetry of the cluster. We call such momentum transfers extinctions and derive an equation whose every zero is an extinction, built from the symmetry species of the two
levels and from the positions and scattering amplitudes of the magnetic sites. The
extinctions follow in closed form in two cases: when each orbit of symmetry-equivalent sites contributes a single occurrence of the mediating species, and when the site phases recur under a subgroup up to one common factor.
\end{abstract}

\maketitle

\section{Introduction}

Inelastic neutron scattering resolves the magnetic transitions of a molecular spin
cluster~\cite{Gudel1977,Furrer1977,Furrer2013}. On a single crystal, present-day
spectrometers record the intensity as a function of the energy
transfer and of all three components of the momentum-transfer vector $\Q$ --
four-dimensional inelastic neutron
scattering~\cite{Baker2012,Garlatti2017,Chiesa2017,Garlatti2020,Garlatti2019}
-- whereas a powder measurement retains only the dependence on the magnitude
of $\Q$. That magnitude alone already carries the distances between the
magnetic sites: for a dimer the powder intensity contains the interference factor $1\pm\sin(QR)/(QR)$, and oscillations of this kind have served to identify transitions in clusters of two and three
sites~\cite{Furrer1979,Guedel1979,Furrer1989}. Structure factors of
small spin-$1/2$ clusters have been computed
exactly~\cite{Haraldsen2005,Haraldsen2016}.

For some transitions the molecular point group fixes the dependence on the
vector $\Q$ completely~\cite{Waldmann2003,Waldmann2007,GT2021a,GT2021b}. This
happens when a single symmetry species mediates the transition and that
species occurs only once among the local spin operators. The intensity is
then a product of a dynamic and a geometric factor. Values of the exchange
and anisotropy parameters determine the dynamic factor and with it the
overall strength of the transition. The geometric factor is a fixed function
of $\Q$, set by the positions and scattering amplitudes of the magnetic sites
and by the point-group symmetry species of the two levels. Functions of this
kind have been tabulated for spin rings and for a series of spin
polyhedra~\cite{GT2021a,GT2021b}. The selection rules and universal $\Q$
dependences that follow from the point group have not yet been examined
experimentally, since single-crystal studies of molecular spin clusters
remain scarce.

For other transitions the condition for a universal $\Q$ dependence fails, in one of two ways: several
species mediate the transition, which happens only when both levels span multidimensional irreducible representations of the point group, or the single mediating species occurs more than
once among the local spin operators. Properties of the wave functions then
enter the intensity, and those depend on the values of the Hamiltonian
parameters. We show here that the point group can still require the intensity
to vanish at particular momentum transfers, for every Hamiltonian that has
the symmetry of the cluster. We call such a momentum transfer an extinction.

From the decomposition of the site amplitudes into irreducible
components under the point group we derive an equation whose every zero
is such an extinction, so that
the two levels enter through their symmetry species and the cluster through
the site positions and the scattering amplitudes. Clusters of different point
groups serve as examples in which the extinction sets are families of
parallel planes, amplitude-dependent curves, lattices of lines or isolated
points, according to the number of conditions that the site phases impose.
Because these extinctions depend on the direction of the momentum transfer,
they are a feature of four-dimensional inelastic neutron scattering. They do not survive a powder average.

\section{Theory}

\subsection{Intensity of a magnetic transition}

For a cluster of $N$ magnetic sites at positions $\Rv_{i}$, each with a spin
operator $\hat{\mathbf{s}}_{i}$, we consider a transition between two different energy levels $n$ and $m$. A level can be degenerate, so $n$ and $m$ label the levels throughout
and $n'$ and $m'$ the states within them. In the customary approximation of a spherical magnetization
density with quenched orbital part around each site, the Fourier transform of the
magnetization density of the cluster is
\begin{equation}
\hat{M}_{\alpha}(\Q)=\sum_{i}F_{i}(Q)\,e^{i\Q\cdot\Rv_{i}}\,\hat{s}_{i\alpha},
\label{eq:fourier}
\end{equation}
where $\alpha$ labels the Cartesian directions and $F_{i}(Q)$ is the scattering
amplitude of site $i$, which collects its magnetic form factor and its $g$ value and
depends on the magnitude of $\Q$ alone.

For a fixed orientation of the cluster we take the intensity of the transition to
be~\cite{Marshall1971}
\begin{equation}
I_{nm}(\Q)=\sum_{n'm'}\sum_{\alpha\beta}
\bigl(\delta_{\alpha\beta}-\hat{Q}_{\alpha}\hat{Q}_{\beta}\bigr)\,
T_{\alpha}^{n'm'}\,T_{\beta}^{n'm'\,*},
\label{eq:intensity}
\end{equation}
where $\hat{\mathbf{Q}}$ is the unit vector along $\Q$ and
$T^{n'm'}_{\alpha}(\Q)=\langle m'|\hat{M}_{\alpha}(\Q)|n'\rangle$.
Equation~(\ref{eq:intensity}) is the cross section up to further factors, among them the
kinematic factor, the square of the magnetic scattering length, the Debye-Waller factor and the thermal population of the initial level, which have no zeros and are therefore
irrelevant for the extinctions.

\subsection{Point-group symmetry}
\label{sec:symmetry}

What follows rests on the symmetry of the spin Hamiltonian and not on its detailed
form. How the point group is represented on the spin states depends on whether the
Hamiltonian is isotropic or anisotropic, the same distinction that underlies the
point-group selection rules and the universal $\Q$ dependences of
Refs.~\onlinecite{GT2021a,GT2021b}. In either case no field is applied and the
Hamiltonian is invariant under time reversal.

An isotropic Hamiltonian has spin-rotational symmetry, so that the total spin $S$ is a
good quantum number. The simplest isotropic spin interaction is the Heisenberg model,
but the treatment applies to every isotropic coupling (biquadratic exchange, multicenter
terms, etc.) that the molecular symmetry admits. The point-group symmetry then acts
through permutations of the site labels.

Each operation $g$ of the molecular point group $G$ maps the set of site positions onto
itself and thereby induces a permutation $\pi_{g}$ of the site
labels~\cite{Waldmann2000}. With $\hat{U}_{g}$ the unitary operator that represents $g$ on the
spin states, the local spin operators transform in the isotropic case as
\begin{equation}
\hat{U}_{g}\,\hat{s}_{i\alpha}\,\hat{U}_{g}^{-1}=\hat{s}_{\pi_{g}(i)\alpha}.
\label{eq:coviso}
\end{equation}

An orbit collects the sites that the permutations interchange, so a cluster whose sites
are all symmetry-equivalent has a single orbit. Note that the map $g\mapsto\pi_{g}$ need not be
one to one. In a cluster whose sites all lie in one plane the reflection in that plane
leaves every site in place, so the permutations form a group smaller than $G$.

Under these permutations the $N$ magnetic sites span a representation of $G$, written
$\Gamma^{(N)}$. One basis vector belongs to each site, and $g$ acts on this
$N$-dimensional space of site amplitudes by the matrix $D(g)$ that takes the basis
vector of site $i$ to that of site $\pi_{g}(i)$, so
$[D(g)\mathbf{v}]_{i}=v_{\pi_{g}^{-1}(i)}$ for a vector $\mathbf{v}$ of that space.

An anisotropic Hamiltonian lacks spin-rotational symmetry, and a permutation of the
site labels alone is in general no symmetry either. A point-group operation remains a
symmetry only as the combination of the two, permuting the site labels and rotating the spin vectors together~\cite{Klemm2008}, and Eq.~(\ref{eq:coviso}) is replaced by
\begin{equation}
\hat{U}_{g}\,\hat{s}_{i\alpha}\,\hat{U}_{g}^{-1}
=\sum_{\beta}\bigl(A_{g}^{-1}\bigr)_{\alpha\beta}\,\hat{s}_{\pi_{g}(i)\beta},
\label{eq:covariance}
\end{equation}
where $A_{g}=\det(R_{g})\,R_{g}$ is the matrix that represents the operation on a spin
vector and $R_{g}$ its matrix in position space. The symmetry group of the anisotropic
case is therefore a subgroup of that of an isotropic Hamiltonian. For half-integral total spin the levels transform according to fermionic
irreducible representations of the double group. The projector construction carries over unchanged, while the time-reversal pairing of Sec.~\ref{sec:species} depends on the type of the representation. No example below uses this case.

Because spin is an axial vector, $A_{g}$ differs from $R_{g}$ only for improper operations, i.e.\ for reflections, inversions and rotoreflections. For a proper
rotation the determinant equals unity and the two matrices coincide, so the spin
vectors are turned about the same axis and by the same angle as the sites. An inversion
leaves the spin vectors unchanged, and a reflection turns them by half a revolution
about the normal of the plane.

The three spin components transform among themselves by the matrices $A_{g}$ of Eq.~(\ref{eq:covariance}). They span a three-dimensional representation of $G$, the
axial-vector representation $\Gamma_{\mathrm{ax}}$, and the local spin operators then
span $\Gamma^{(3N)}=\Gamma^{(N)}\otimes\Gamma_{\mathrm{ax}}$.

\subsection{Transition-mediating species}
\label{sec:species}

The decompositions of $\Gamma^{(N)}$ and $\Gamma^{(3N)}$ into irreducible components
give the site species and the species of the local spin operators, respectively. An
irrep $\Gamma_{l}$ can occur more than once. Its occurrences together span the channel of
$\Gamma_{l}$. The two levels of a transition transform according to the level species
$\Gamma_{n}$ and $\Gamma_{m}$, and a channel mediates the transition when $\Gamma_{l}$
is contained in $\Gamma_{n}^{*}\otimes\Gamma_{m}$.

The effective multiplicity $a_{l}$ counts the occurrences of $\Gamma_{l}$, with one
exception: in the isotropic description one occurrence of the totally symmetric
species $\Gamma_{1}$ is discarded, namely the uniform amplitude vector, whose operator is the total spin $\hat{\mathbf{S}}=\sum_{i}\hat{\mathbf{s}}_{i}$, which commutes with the Hamiltonian and does not connect different levels. In the
anisotropic description the total spin is not conserved and nothing is discarded. The set $\mathcal{A}_{nm}$ collects the species with $a_{l}\geq1$ for which
$\Gamma_{n}^{*}\otimes\Gamma_{m}$ contains $\Gamma_{l}$.

Time reversal, under which the Hamiltonian is invariant, makes a level whose
species has complex characters degenerate with a level of the conjugate species. The level species and the entries of $\mathcal{A}_{nm}$ are then the
paired sums $\Gamma\oplus\Gamma^{*}$. Every point group of the worked examples below has only real characters.

An example is the ring of eight sites of Sec.~\ref{sec:ring}, in which two kinds of
magnetic site alternate. Its molecular point group is $D_{4h}$, but the ring is planar,
so $\sigma_{h}$ leaves every site in place and the permutations form $D_{4}$. In the
isotropic description the sites span $\Gamma^{(N)}=2A_{1}+B_{1}+B_{2}+2E$, and the
effective multiplicities are $a_{A_{1}}=1$, the uniform vector being discarded from the
two occurrences, $a_{B_{1}}=a_{B_{2}}=1$, and $a_{E}=2$. A transition between two levels of
species $E$ has $E\otimes E=A_{1}+A_{2}+B_{1}+B_{2}$, of which $A_{2}$ is not a site
species, so $\mathcal{A}_{nm}=\{A_{1},B_{1},B_{2}\}$.

In the anisotropic description the spin rotations make $D_{4h}$ act faithfully. The
sites span $2A_{1g}+B_{1g}+B_{2g}+2E_{u}$, and with $\Gamma_{\mathrm{ax}}=A_{2g}+E_{g}$
the local spin operators span
$\Gamma^{(3N)}=2A_{2g}+B_{1g}+B_{2g}+4E_{g}+2A_{1u}+2A_{2u}+2B_{1u}+2B_{2u}+2E_{u}$. A transition between two
levels of species $E_{u}$ then has $\mathcal{A}_{nm}=\{A_{2g},B_{1g},B_{2g}\}$.

\subsection{Channel amplitudes}

The amplitude vector $\mathbf{c}(\Q)$ collects the coefficients with which the local
spin operators enter Eq.~(\ref{eq:fourier}). Its component at site $i$ is
\begin{equation}
c_{i}(\Q)=F_{i}(Q)\,e^{i\Q\cdot\Rv_{i}},
\label{eq:cvec}
\end{equation}
and it lies in the $N$-dimensional space on which the matrices $D(g)$ act. Since
$R_{g}\Rv_{i}=\Rv_{\pi_{g}(i)}$ and the sites of one orbit share one scattering
amplitude, replacing $\Q$ by the turned momentum transfer $R_{g}\Q$ permutes the
components,
\begin{equation}
D(g)\,\mathbf{c}(\Q)=\mathbf{c}(R_{g}\Q).
\label{eq:cturn}
\end{equation}

For an isotropic Hamiltonian, let $P_{\Gamma_{l}}$ project an amplitude vector onto the
channel of $\Gamma_{l}$. The projectors add up to the identity, so
$\mathbf{c}(\Q)=\sum_{l}P_{\Gamma_{l}}\mathbf{c}(\Q)$. By Eq.~(\ref{eq:coviso}), an
operator built from a coefficient vector $\mathbf{v}$ transforms as the vector itself,
\begin{equation}
\hat{U}_{g}\Bigl(\sum_{i}v_{i}\,\hat{s}_{i\alpha}\Bigr)\hat{U}_{g}^{-1}
=\sum_{i}v_{i}\,\hat{s}_{\pi_{g}(i)\alpha}
=\sum_{j}\bigl(D(g)\mathbf{v}\bigr)_{j}\,\hat{s}_{j\alpha},
\label{eq:intertwine}
\end{equation}
so the part of $\hat{M}_{\alpha}(\Q)$ transforming as $\Gamma_{l}$ is
\begin{equation}
\hat{M}^{(l)}_{\alpha}(\Q)=\sum_{i}\bigl(P_{\Gamma_{l}}\mathbf{c}(\Q)\bigr)_{i}\,
\hat{s}_{i\alpha}.
\label{eq:split}
\end{equation}

Let the channel of $\Gamma_{l}$ have effective multiplicity one, $a_{l}=1$, the case
from which the universal $\Q$ dependences arise. It has the dimension $d_{l}$ of $\Gamma_{l}$, with an orthonormal basis $\mathbf{u}_{1},\dots,\mathbf{u}_{d_{l}}$. Writing
$P_{\Gamma_{l}}\mathbf{c}(\Q)=\sum_{q}w_{q}(\Q)\,\mathbf{u}_{q}$,
Eq.~(\ref{eq:split}) reads
$\hat{M}^{(l)}_{\alpha}(\Q)=\sum_{q}w_{q}(\Q)\,\hat{O}_{q\alpha}$ with
$\hat{O}_{q\alpha}=\sum_{i}u_{qi}\,\hat{s}_{i\alpha}$. The $d_{l}$ operators form one
irreducible set, and by the Wigner-Eckart theorem for the point
group~\cite{Altmann1994} their matrix elements are
\begin{equation}
\langle m'|\hat{O}_{q\alpha}|n'\rangle=r_{nm}\,C^{\,n'm'}_{q\alpha}.
\label{eq:we}
\end{equation}
The coefficients $C$ are fixed by symmetry and proportional to Clebsch--Gordan
coefficients of the point group, with their normalization chosen so that
Eq.~(\ref{eq:factor}) below holds. For the simply reducible point groups of this
work a single reduced matrix element $r_{nm}$ appears per occurrence, and it
carries the dynamics of the transition. In the isotropic description the levels transform under $SU(2)\times G$ and the
operators $\hat{O}_{q\alpha}$ as $(1,\Gamma_{l})$, so that $C$ contains a spin as well as a point-group Clebsch--Gordan coefficient, and the spin selection rule $\lvert S_{m}-S_{n}\rvert\leq1\leq S_{m}+S_{n}$ holds.

The channel amplitude
\begin{equation}
\Phi_{l}(\Q)=\bigl\lVert P_{\Gamma_{l}}\mathbf{c}(\Q)\bigr\rVert
\label{eq:chanamp}
\end{equation}
measures the part of the amplitude vector that lies in the channel, and its square is the channel weight. In the isotropic description, where one occurrence of the totally symmetric
species $\Gamma_{1}$ is discarded, its amplitude is computed with the reduced
projector $P_{\Gamma_{1}}-\mathbf{1}\mathbf{1}^{\mathsf{T}}/N$, with $\mathbf{1}$
the uniform vector. When only a channel with
$a_{l}=1$ mediates the transition, Eqs.~(\ref{eq:split}) and (\ref{eq:we}) give
\begin{equation}
I_{nm}(\Q)=\lvert r_{nm}\rvert^{2}\,\Phi_{l}(\Q)^{2},
\label{eq:factor}
\end{equation}
where $\lvert r_{nm}\rvert^{2}$ is the dynamic factor of the transition and
$\Phi_{l}(\Q)^{2}$ its geometric factor. Spin-rotational symmetry makes the tensor
$\sum_{n'm'}T_{\alpha}T_{\beta}^{*}$
proportional to $\delta_{\alpha\beta}$, so the transverse projector of
Eq.~(\ref{eq:intensity}) contributes its trace, a factor of two, at every momentum transfer. This factor is absorbed in the normalization of the coefficients $C$.

A transition of this kind has a universal $\Q$ dependence: the projection of Eq.~(\ref{eq:chanamp}) reproduces the tabulated
functions of Refs.~\onlinecite{GT2021a,GT2021b}, up to a constant factor per channel.

A channel with $a_{l}\geq2$ holds $a_{l}$ occurrences of $\Gamma_{l}$, each with its own set
of operators and its own reduced matrix element $r^{(1)}_{nm},\dots,r^{(a_{l})}_{nm}$.
The matrix elements of $\hat{M}^{(l)}_{\alpha}(\Q)$ are then combinations of these
numbers, so how the intensity varies with $\Q$ depends on the dynamics, and thus on the
Hamiltonian parameters. Where $P_{\Gamma_{l}}\mathbf{c}(\Q)$ vanishes for every
mediating species, however, the intensity is zero independent of the values of the reduced matrix elements. Those momentum transfers are the subject of this work.

\subsection{The extinction indicator}
\label{sec:indicator}

If $P_{\Gamma_{l}}\mathbf{c}(\Q)$ is the zero vector, $\hat{M}^{(l)}_{\alpha}(\Q)$ of Eq.~(\ref{eq:split}) is the zero operator. At a
momentum transfer where this happens for every species of $\mathcal{A}_{nm}$, the
intensity vanishes and the transition is called dark. Adding the squared lengths of
the projections combines these conditions, one per mediating species, into a single non-negative number -- the extinction indicator -- that vanishes exactly where all of them are met.

In the isotropic description the indicator is the sum of squared norms
\begin{equation}
\mathcal{E}^{\mathrm{iso}}_{nm}(\Q)=\sum_{\Gamma_{l}\in\mathcal{A}_{nm}}
\Bigl\lVert\Bigl(P_{\Gamma_{l}}
-\delta_{l,1}\,\frac{\mathbf{1}\mathbf{1}^{\mathsf{T}}}{N}\Bigr)\mathbf{c}(\Q)\Bigr\rVert^{2},
\label{eq:indicator}
\end{equation}
where $\mathbf{1}\mathbf{1}^{\mathsf{T}}/N$ projects onto the uniform vector and
$\delta_{l,1}$ is unity for $\Gamma_{1}$ and zero otherwise. The subtraction removes the occurrence that the effective multiplicity does not count.

Equation~(\ref{eq:indicator}) requires a projection at every momentum transfer.
Written as a sum over the group, the projector separates the level species from the
geometry,
\begin{equation}
P_{\Gamma_{l}}=\frac{d_{l}}{|G|}\sum_{g\in G}\chi_{l}(g)^{*}\,\mathcal{D}(g),
\label{eq:proj}
\end{equation}
with $\chi_{l}$ the character of $\Gamma_{l}$. Here $\mathcal{D}(g)=D(g)$, and in
the anisotropic description below $\mathcal{D}(g)=D(g)\otimes A_{g}$. Since
$\lVert P_{\Gamma_{l}}\mathbf{c}\rVert^{2}=\langle\mathbf{c},P_{\Gamma_{l}}\mathbf{c}\rangle$,
only the numbers $\langle\mathbf{c},D(g)\mathbf{c}\rangle$ are needed, and because
the sites of one orbit share one scattering amplitude,
\begin{equation}
\langle\mathbf{c},D(g)\mathbf{c}\rangle=S_{g}(\Q)=\sum_{i}F_{i}(Q)^{2}\,
e^{i\Q\cdot(\Rv_{i}-\Rv_{\pi_{g}(i)})}.
\label{eq:structure}
\end{equation}
The two levels enter through the characters,
\begin{equation}
\Psi_{nm}(g)=\sum_{\Gamma_{l}\in\mathcal{A}_{nm}}d_{l}\,\chi_{l}(g),
\label{eq:species}
\end{equation}
and the isotropic indicator takes the closed form
\begin{equation}
\mathcal{E}^{\mathrm{iso}}_{nm}(\Q)=\frac{1}{|G|}\sum_{g\in G}
\Psi_{nm}(g)^{*}\,S_{g}(\Q)
-\delta^{\Gamma_{1}}_{nm}\,\frac{\bigl|\mathbf{1}^{\mathsf{T}}\mathbf{c}(\Q)\bigr|^{2}}{N},
\label{eq:masteriso}
\end{equation}
where $\delta^{\Gamma_{1}}_{nm}$ equals unity when $\Gamma_{1}$ belongs to
$\mathcal{A}_{nm}$ and zero otherwise. The second term is the subtraction of
Eq.~(\ref{eq:indicator}). The transverse projector of Eq.~(\ref{eq:intensity}), written out in Eq.~(\ref{eq:transproj}) below, contributes only the constant factor absorbed in the normalization of the
coefficients $C$ and leaves no trace here.

In the anisotropic description the projection is taken in the $3N$-dimensional
space of $\Gamma^{(3N)}$, the tensor product of the space of site amplitudes with
the three-dimensional space of spin directions. A product $\mathbf{a}\otimes\hat{\mathbf{e}}$ stands for the operator
$\sum_{i}a_{i}\,\hat{\mathbf{s}}_{i}\cdot\hat{\mathbf{e}}$. Which operators to
project follows from the transverse projector $\Pi$ of Eq.~(\ref{eq:transproj}):
with $\hat{\mathbf{e}}_{1}$ and $\hat{\mathbf{e}}_{2}$ two unit vectors that
complete $\hat{\mathbf{Q}}$ to an orthonormal basis,
\begin{equation}
\Pi_{\alpha\beta}=\delta_{\alpha\beta}-\hat{Q}_{\alpha}\hat{Q}_{\beta}
=\sum_{\sigma=1,2}\hat{e}_{\sigma\alpha}\,\hat{e}_{\sigma\beta},
\label{eq:transproj}
\end{equation}
and the intensity becomes
\begin{equation}
I_{nm}(\Q)=\sum_{n'm'}\sum_{\sigma=1,2}
\Bigl|\sum_{\alpha}\hat{e}_{\sigma\alpha}\,T^{n'm'}_{\alpha}(\Q)\Bigr|^{2}.
\label{eq:transverse}
\end{equation}
The operator behind each term is the transverse spin component
$\sum_{\alpha}\hat{e}_{\sigma\alpha}\hat{M}_{\alpha}(\Q)
=\sum_{i}c_{i}(\Q)\,\hat{\mathbf{s}}_{i}\cdot\hat{\mathbf{e}}_{\sigma}$, whose
vector is $\mathbf{c}(\Q)\otimes\hat{\mathbf{e}}_{\sigma}$. Projecting it gives
the weight that the channel of $\Gamma$ carries at $\Q$,
\begin{equation}
\mathcal{K}_{\Gamma}(\Q)=\sum_{\sigma=1,2}\bigl\lVert P_{\Gamma}
\bigl[\mathbf{c}(\Q)\otimes\hat{\mathbf{e}}_{\sigma}\bigr]\bigr\rVert^{2},
\label{eq:weight}
\end{equation}
the anisotropic counterpart of $\Phi_{l}(\Q)^{2}$. The sums over $\sigma$ are independent of the choice of $\hat{\mathbf{e}}_{1}$ and $\hat{\mathbf{e}}_{2}$.
The extinction indicator is the sum over the mediating species,
\begin{equation}
\mathcal{E}^{\mathrm{aniso}}_{nm}(\Q)=\sum_{\Gamma_{l}\in\mathcal{A}_{nm}}
\mathcal{K}_{\Gamma_{l}}(\Q).
\label{eq:indicataniso}
\end{equation}
When only a channel with $a_{l}=1$ mediates the transition,
$I_{nm}(\Q)=\lvert r_{nm}\rvert^{2}\mathcal{K}_{\Gamma_{l}}(\Q)$, the counterpart
of Eq.~(\ref{eq:factor}).

The corresponding number of Eq.~(\ref{eq:structure}) factorizes,
\begin{equation}
\bigl\langle\mathbf{c}\otimes\hat{\mathbf{e}}_{\sigma},
(D(g)\otimes A_{g})(\mathbf{c}\otimes\hat{\mathbf{e}}_{\sigma})\bigr\rangle
=S_{g}(\Q)\,\hat{\mathbf{e}}_{\sigma}^{\mathsf{T}}A_{g}\hat{\mathbf{e}}_{\sigma},
\label{eq:factorize}
\end{equation}
and the second factor, summed over the two directions, is the trace of the product
of $A_{g}$ with the transverse projector: with $A_{g}=\det(R_{g})R_{g}$,
\begin{equation}
\begin{split}
\chi_{\perp}(g,\hat{\mathbf{Q}})&=\sum_{\sigma}\hat{\mathbf{e}}_{\sigma}^{\mathsf{T}}A_{g}\hat{\mathbf{e}}_{\sigma}
=\operatorname{Tr}\bigl(\Pi A_{g}\bigr)\\
&=\det(R_{g})\bigl[\operatorname{Tr}R_{g}-\hat{\mathbf{Q}}\cdot R_{g}\hat{\mathbf{Q}}\bigr].
\end{split}
\label{eq:chiperp}
\end{equation}
In closed form, the anisotropic indicator is
\begin{equation}
\mathcal{E}^{\mathrm{aniso}}_{nm}(\Q)=\frac{1}{|G|}\sum_{g\in G}
\Psi_{nm}(g)^{*}\;\chi_{\perp}(g,\hat{\mathbf{Q}})\;S_{g}(\Q).
\label{eq:master}
\end{equation}
Both indicators state the same fact: at their zeros the geometry closes every mediating channel. Parameter
independence refers to the exchange and anisotropy parameters. The site positions and scattering amplitudes are fixed by the structure of the cluster. Every
extinction reported below is a zero of Eq.~(\ref{eq:masteriso}) or of
Eq.~(\ref{eq:master}), according to the description.

\subsection{Extinctions in closed form}
\label{sec:closedforms}

This subsection gives the zeros of the indicator without a projection at
every momentum transfer. A sufficient character condition covers the momentum
transfers on symmetry elements, and in two situations the zero set follows in
closed form, as an explicit equation. The character condition uses the stabilizer of a nonzero momentum transfer, the subgroup
$H(\Q)=\{g\in G:R_{g}\Q=\Q\}$. By Eq.~(\ref{eq:cturn}) the vector
$\mathbf{c}(\Q)$ is invariant under $H(\Q)$.

In the isotropic description a matrix element of $\hat{M}^{(l)}_{\alpha}(\Q)$
between the two levels is nonzero only if the totally symmetric species of
$H(\Q)$ occurs in $\Gamma_{n}^{*}\otimes\Gamma_{m}$ restricted to $H(\Q)$. The number of occurrences is
\begin{equation}
\nu^{\mathrm{iso}}_{nm}(\Q)=\frac{1}{|H(\Q)|}\sum_{g\in H(\Q)}
\chi_{m}(g)^{*}\,\chi_{n}(g).
\label{eq:nuiso}
\end{equation}
In the anisotropic description the transverse projector enters: an operation of
$H(\Q)$ maps the plane perpendicular to $\Q$ onto itself, so the two transverse
components of $\hat{M}_{\alpha}(\Q)$ transform among themselves and span a
two-dimensional representation $\Gamma_{\perp}$ of $H(\Q)$ with character
$\chi_{\perp}(g)=\det(R_{g})[\operatorname{Tr}R_{g}-1]$, which is
Eq.~(\ref{eq:chiperp}) at an operation with $R_{g}\Q=\Q$. The count becomes
\begin{equation}
\nu_{nm}(\Q)=\frac{1}{|H(\Q)|}\sum_{g\in H(\Q)}
\chi_{m}(g)^{*}\,\chi_{\perp}(g)\,\chi_{n}(g).
\label{eq:nu}
\end{equation}

A transition is dark wherever $\nu_{nm}$ vanishes, and since $\nu_{nm}$ depends on $\Q$
only through the stabilizer, it is dark at every momentum transfer with that stabilizer.
That $\nu_{nm}=0$ closes every mediating channel, and thus sets the
extinction indicator to zero, follows from decomposing $\Gamma_{n}^{*}\otimes\Gamma_{m}$ into species of $G$. Each species enters
Eq.~(\ref{eq:nu}) with the number of times $\Gamma_{\perp}$ occurs in its restriction,
and these numbers are non-negative. By Eq.~(\ref{eq:cturn}) the vector $\mathbf{c}(\Q)$
is invariant under $H(\Q)$, so the projected vector transforms as $\Gamma_{\perp}$ and
has no part in a channel whose restriction shares no irreducible constituent with $\Gamma_{\perp}$. The converse fails:
the projection can vanish where $\nu_{nm}$ does not, so Eq.~(\ref{eq:nu}) is the weaker
of the two conditions.

For a trivial stabilizer Eqs.~(\ref{eq:nuiso}) and (\ref{eq:nu}) give
$\nu^{\mathrm{iso}}_{nm}=d_{n}d_{m}$ and $\nu_{nm}=2d_{n}d_{m}$, so a vanishing $\nu_{nm}$
requires a nontrivial one. An operation other than the identity fixes $\Q$ only if it is
a rotation about an axis along $\Q$ or a reflection in a plane containing $\Q$. Equation~(\ref{eq:nu}) therefore vanishes only on rotation axes and in mirror planes of the cluster.

The first of the two closed forms concerns the channels to which every orbit of sites
contributes one occurrence. Let $t_{B}(\Q)=\sum_{i\in B}c_{i}(\Q)$ be the amplitude sum over an orbit $B$ of
$\lvert B\rvert$ sites and $\bar{t}_{B}=t_{B}/\lvert B\rvert$ the orbit mean. Every
orbit contributes one occurrence of the totally symmetric species, whose effective
multiplicity in the isotropic description is thus the number of orbits less one, and the term that this channel
contributes to Eq.~(\ref{eq:indicator}) is
\begin{equation}
\Bigl\lVert\Bigl(P_{\Gamma_{1}}-\frac{\mathbf{1}\mathbf{1}^{\mathsf{T}}}{N}\Bigr)
\mathbf{c}(\Q)\Bigr\rVert^{2}
=\sum_{B}\frac{\lvert t_{B}\rvert^{2}}{\lvert B\rvert}
-\frac{1}{N}\Bigl\lvert\sum_{B}t_{B}\Bigr\rvert^{2}.
\label{eq:orbitiso}
\end{equation}
It vanishes exactly where all orbits share one value of $\bar{t}_{B}$.

In a point group with a principal axis the spin component along the axis spans a
one-dimensional species $\Gamma_{z}$, and for each orbit the operator
\begin{equation}
\hat{S}_{Bz}=\sum_{i\in B}\hat{s}_{iz}
\label{eq:orbitspin}
\end{equation}
transforms as $\Gamma_{z}$, since a symmetry operation maps the orbit onto itself and
multiplies the axial spin component by the character of $\Gamma_{z}$. When every site lies on the axis or in a horizontal mirror plane of the
group, these operators are all the occurrences of $\Gamma_{z}$, one per orbit. Otherwise the spin components perpendicular to the axis contribute further
occurrences, and the right-hand side of Eq.~(\ref{eq:orbitaniso}) below is then a lower
bound of the channel weight rather than the weight itself. Their vectors $\mathbf{1}_{B}\otimes\hat{\mathbf{z}}$,
with $\mathbf{1}_{B}$ equal to one on the sites of $B$ and zero elsewhere, are mutually
orthogonal, have squared length $\lvert B\rvert$, and span these occurrences. Under the preceding condition, projecting
$\mathbf{c}(\Q)\otimes\hat{\mathbf{e}}_{\sigma}$ onto them and adding the squared lengths for the
two directions, with $\sum_{\sigma}(\hat{\mathbf{z}}\cdot\hat{\mathbf{e}}_{\sigma})^{2}=\sin^{2}\theta_{\Q}$
and $\theta_{\Q}$ the angle between $\Q$ and the axis, gives
\begin{equation}
\mathcal{K}_{\Gamma_{z}}(\Q)=\sin^{2}\theta_{\Q}
\sum_{B}\frac{\lvert t_{B}\rvert^{2}}{\lvert B\rvert}.
\label{eq:orbitaniso}
\end{equation}
It vanishes on the axis, where $\sin\theta_{\Q}=0$, and elsewhere only where every orbit
sum $t_{B}$ vanishes.

The second closed form is not tied to the orbits. It concerns the momentum transfers at
which the amplitude vector is a common eigenvector of the matrices of a subgroup $K$ of
$G$,
\begin{equation}
D(g)\,\mathbf{c}(\Q)=\omega(g)\,\mathbf{c}(\Q),\qquad g\in K.
\label{eq:eigen}
\end{equation}
Applying two operations in turn shows that $\omega$ is a one-dimensional representation
of $K$, and $\lvert\omega(g)\rvert=1$ because $D(g)$ preserves lengths. By
Eq.~(\ref{eq:cturn}) the left-hand side of Eq.~(\ref{eq:eigen}) is $\mathbf{c}(R_{g}\Q)$,
so the condition reads $\mathbf{c}(R_{g}\Q)=\omega(g)\mathbf{c}(\Q)$ and does not require
$g$ to fix $\Q$. Every operation of the stabilizer satisfies it with $\omega=1$, so
$H(\Q)\subseteq K$ and the axes and planes of Eq.~(\ref{eq:nu}) are among the solutions. But $K$ can contain operations that turn $\Q$ into another momentum transfer and can be
nontrivial where $H(\Q)$ is not, and Eq.~(\ref{eq:nu}), built from the stabilizer alone,
does not reach the momentum transfers found in this way.

In components, Eq.~(\ref{eq:eigen}) with Eq.~(\ref{eq:cvec}), whose amplitudes agree
within an orbit, gives the congruence
\begin{equation}
\Q\cdot\bigl(\Rv_{\pi_{g}(i)}-\Rv_{i}\bigr)\equiv-\arg\omega(g)
\quad(\mathrm{mod}\;2\pi)
\label{eq:congruence}
\end{equation}
for every $g$ in $K$ and every site $i$. A site that $g$ leaves in place has
$\mathbf{a}=\Rv_{\pi_{g}(i)}-\Rv_{i}=\mathbf{0}$ and admits $\omega(g)=1$ alone. For
$\mathbf{a}\neq\mathbf{0}$ the solutions of one congruence are
\begin{equation}
\Q=\frac{b+2\pi n}{\lvert\mathbf{a}\rvert^{2}}\,\mathbf{a}+\Q_{\perp},
\qquad n\in\mathbb{Z},\quad \mathbf{a}\cdot\Q_{\perp}=0,
\label{eq:planes}
\end{equation}
with $b=-\arg\omega(g)$, a family of parallel planes with normal $\mathbf{a}$, spaced
$2\pi/\lvert\mathbf{a}\rvert$ apart. Taken together, the congruences fix the component
of $\Q$ in the space spanned by the difference vectors and leave the orthogonal
component free: the solutions are parallel planes when the difference vectors span one
direction, parallel lines when they span two, and isolated points when they span all three. The simplest case is a pair of
equivalent sites: $K=\{E,C_{2}\}$ with $\omega=1$ gives the planes on which the
interference factor $1-\cos[\Q\cdot(\Rv_{2}-\Rv_{1})]$ of a spin dimer
vanishes~\cite{Furrer1979}. Two difference vectors $\mathbf{a}$ and $\lambda\mathbf{a}$ along one
direction impose two families of Eq.~(\ref{eq:planes}) at once. For rational $\lambda$ they share a family of planes whenever the two congruences are compatible, and
otherwise share none, for irrational $\lambda$ and $\omega=1$ only the plane through
the origin.

At these momentum transfers $\mathbf{c}(\Q)$ lies in the subspace on which $K$ acts
through $\omega$. Since $P_{\Gamma_{l}}$ commutes with the operations,
$P_{\Gamma_{l}}\mathbf{c}(\Q)$ obeys Eq.~(\ref{eq:eigen}) as well and lies in the
channel of $\Gamma_{l}$, on which $K$ acts as the restriction of $\Gamma_{l}$. If that restriction does not contain $\omega$, the projection is the zero vector and the channel
amplitude $\Phi_{l}(\Q)$ of Eq.~(\ref{eq:chanamp}) vanishes. In the anisotropic description a channel $\Gamma_{l}$ receives contributions from
the site species $\Gamma_{\mu}$ with
$\Gamma_{l}\subset\Gamma_{\mu}\otimes\Gamma_{\mathrm{ax}}$, and its weight
$\mathcal{K}_{\Gamma_{l}}$ vanishes wherever the full projections
$\lVert P_{\Gamma_{\mu}}\mathbf{c}(\Q)\rVert$ vanish for all of them.

The dimension of an extinction set follows from counting conditions. In the isotropic description a channel of effective multiplicity
$a_{l}$ whose species has dimension $d_{l}$ imposes up to
$2a_{l}d_{l}$ real conditions, one complex condition per component of
the projected vector. In the anisotropic description both transverse
directions must close, which doubles that number. Inversion symmetry,
time reversal and shared factors lower these counts. At a centrosymmetric cluster a
one-dimensional species of effective multiplicity one leaves a single real
condition, so that its zero set is a surface -- in the anisotropic
description provided its occurrence lies in a one-dimensional spin
species. That is always the case for the spin component along the
principal axis, and also for the perpendicular components when these
span two one-dimensional species. Only when the perpendicular
components span a two-dimensional species does an occurrence there
require the projected vector to be parallel to $\Q$ and leave two
conditions. A
two-dimensional species leaves two conditions as well, so that its zero set
is a curve. When the effective multiplicity is two or more, the indicator is
a sum of several squared amplitudes, and their common zero drops to a curve
or to isolated points unless they share a factor. Families of parallel planes
are the exception: the congruences of Eq.~(\ref{eq:congruence}) impose a
single linear condition that sets every squared amplitude to zero, so that a
two-dimensional set is possible.

\section{Results and discussion}
\label{sec:results}

\begin{figure*}[t]
\centering
\includegraphics[width=0.82\textwidth]{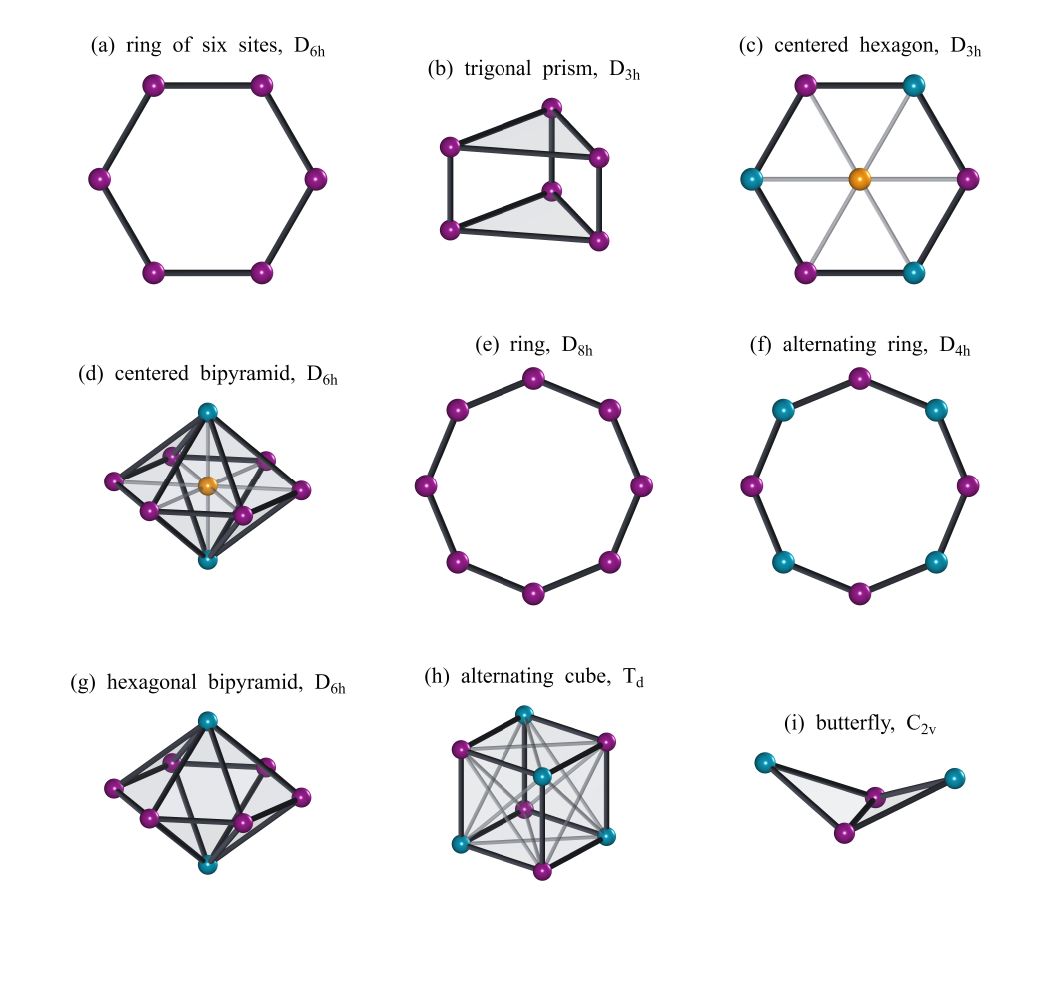}
\caption{\label{fig:cluster}Clusters treated in this work: (a) ring of six
sites of Sec.~\ref{sec:hexring}, (b) trigonal prism of Sec.~\ref{sec:prism},
(c) centered hexagon and (d) centered bipyramid of Sec.~\ref{sec:zbipy},
(e) ring of eight equal sites and (f) ring of two alternating sites of
Sec.~\ref{sec:ring}, (g) hexagonal bipyramid of Sec.~\ref{sec:bipy},
(h) alternating cube of the Supplemental Material~\cite{suppmat}, and
(i) butterfly arrangement, also of Sec.~\ref{sec:ring}. Colors distinguish
orbits of symmetry-equivalent sites. Dark sticks are the bonds that define
the shape, light sticks the links to the central site in (c) and (d) and
within each tetrahedron in (h).}
\end{figure*}

This section illustrates the theory on a sequence of clusters: a ring of six
equal sites, a trigonal prism, a centered hexagon and a centered bipyramid, a ring of two alternating sites together with a butterfly arrangement, and a hexagonal bipyramid, whose invariant Hamiltonians are stated in their subsections. An alternating cube, whose
extinction set consists of isolated points, is treated in the Supplemental
Material. Table~\ref{tab:overview} summarizes all examples, and
Fig.~\ref{fig:cluster} shows the clusters in real space. The Supplemental Material~\cite{suppmat} also derives the power law that governs the rise of the intensity beside an extinction on a rotation axis.

\begin{table*}[t]
\caption{\label{tab:overview}The clusters of this section and of the
Supplemental Material, the mechanism that closes their mediating channels, the dimension of the extinction set, and the aspect that each example illustrates. The ring of eight equal
sites of Sec.~\ref{sec:ring} has universal functions only and is not
listed.}
\begin{ruledtabular}
\begin{tabular}{llll}
Cluster & Mechanism & Set & Aspect \\ \colrule
Ring of six & recurrence, $C_{2}$ & lines & benchmark; two channels \\
Trigonal prism & recurrence, $\sigma_{h}$ and $C_{3}$ & planes, lines & complementary families \\
Centered hexagon & recurrence ($C_{3}$); orbit means & lines; none & existence set by the amplitudes \\
Centered bipyramid & orbit means balance & curves or none & position and existence set by the amplitudes \\
Alternating cube & recurrence spans $\mathbb{R}^{3}$ & points, lines & trivial
stabilizers (Supplemental Material) \\
Alternating ring & orbit sums; irrationality & axis, lines & extinction vs.\ accidental zeros \\
Butterfly & recurrence, $K=G$ & lines & congruences alone suffice \\
Hexagonal bipyramid & orbit sums; full Eq.~(\ref{eq:master}) & curves, arcs & zeros beyond closed forms \\
\end{tabular}
\end{ruledtabular}
\end{table*}

The intensity maps are computed by exact diagonalization, with $s=1/2$ at
every site, and the predicted sets follow from Eq.~(\ref{eq:master}) or
(\ref{eq:masteriso}), the positions of the magnetic sites and their
scattering amplitudes. Each cluster is placed with its principal axis along
$z$ and with one magnetic site on the positive $x$ axis. The cube of the Supplemental Material is placed by the coordinates stated there. The
orientation is part of the specification, because the maps below are sections
through planes in $\Q$ space. All predicted extinction sets below were
verified by exact diagonalization. Site positions are dimensionless. Each intensity map is normalized to its largest value. The color scale of each map, linear or logarithmic, is shown beside it, and values below the lower end of a logarithmic scale are drawn at that value.

Every Hamiltonian below is bilinear in the spins,
\begin{equation}
\mathcal{H}=\sum_{i<j}\hat{\mathbf{s}}_{i}^{\mathsf{T}}\,
\mathsf{J}_{ij}\,\hat{\mathbf{s}}_{j},
\label{eq:model}
\end{equation}
with $\mathsf{J}_{ij}$ a real three-by-three matrix on the bond between the sites $i$ and $j$. The isotropic models keep the scalar coupling of each
bond, $\mathsf{J}_{ij}=J_{ij}\openone$. Couplings are stated per orbit of
symmetry-equivalent bonds. In the anisotropic examples the tensor of one representative bond is given. It is invariant under the operations that fix
that bond, and the tensors of the other bonds follow by symmetry.

\begin{figure*}[t]
\includegraphics[width=\textwidth]{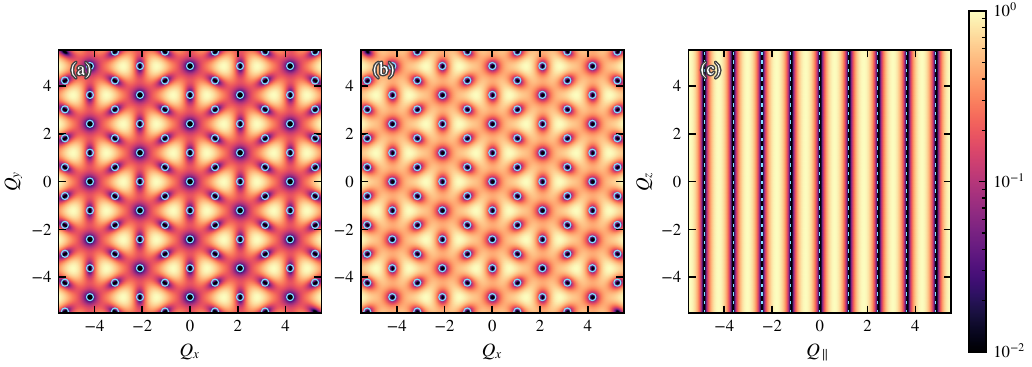}
\caption{\label{fig:hexring}Computed intensity of two $E_{1}$--$E_{2}$
transitions of the ring of six equal sites specified in the text, from the
$E_{1}$ level at $E=-1$ ($S=1$) to the two lowest $E_{2}$ levels. Panel (a)
is the transition to the $E_{2}$ level at $E=-1.2808$ ($S=1$), panel (b) that
to the $E_{2}$ level at $E=0$ ($S=2$), both in the plane $Q_{z}=0$ (two opposite sites lie on the $x$ axis). Panel (c) shows the transition of panel
(a) in the vertical plane that contains the sixfold axis and a shortest
translation vector of the triangular lattice of Eq.~(\ref{eq:hexlattice}),
so that it cuts a row of the extinction lines. Circles and dashed lines mark the extinction lattice of
Eq.~(\ref{eq:hexlattice}).}
\end{figure*}

\subsection{Ring of six sites: extinctions on a lattice of lines}
\label{sec:hexring}

The first example is the isotropic $D_{6}$ spin
ring. All sites are symmetry-equivalent and lie in the $xy$ plane -- the
point group is $D_{6h}$, and the permutations form $D_{6}$
(Sec.~\ref{sec:symmetry}) -- and they span
$\Gamma^{(N)}=A_{1}+B_{1}+E_{1}+E_{2}$. The $A_{1}$ operator corresponds to the
total spin $\hat{\mathbf{S}}$, which is discarded, so that the channels are
$B_{1}$, $E_{1}$ and $E_{2}$. Every pair of level species except $(E_{1},E_{2})$
admits at most one of these channels, and each transition so mediated has a
universal $\Q$ dependence, the case of Eq.~(\ref{eq:factor}). For these transitions Eq.~(\ref{eq:masteriso}) returns the tabulated
functions of Ref.~\onlinecite{GT2021a}, and the extinctions are their zeros.

The product
$E_{1}\otimes E_{2}=B_{1}+B_{2}+E_{1}$ contains the site species $B_{1}$ and $E_{1}$,
so that no universal function exists~\cite{GT2021a}. We define the cyclic ring sums of the six sites, numbered
$j=0,\dots,5$, as
\begin{equation}
t_{k}(\Q)=\sum_{j=0}^{5}e^{-2\pi ikj/6}\,c_{j}(\Q),
\label{eq:hexsums}
\end{equation}
with all scattering amplitudes set to 1, so that
$c_{j}(\Q)=e^{i\Q\cdot\Rv_{j}}$. The two channel weights are
$\Phi_{B_{1}}^{2}=\lvert t_{3}\rvert^{2}/6$ and
$\Phi_{E_{1}}^{2}=(\lvert t_{1}\rvert^{2}+\lvert t_{5}\rvert^{2})/6$. Together they
vanish exactly where $\sin(\Q\cdot\Rv_{j})=0$ at every site,
\begin{equation}
\Q\cdot\Rv_{j}\in\pi\mathbb{Z},\qquad j=0,\dots,5.
\label{eq:hexlattice}
\end{equation}
This is Eq.~(\ref{eq:congruence}) with the subgroup $K=\{E,C_{2}\}$ and $\omega=1$: the
twofold rotation about the principal axis carries each site to the opposite one, the difference vectors are $-2\Rv_{i}$, and at the momentum transfers of Eq.~(\ref{eq:hexlattice}) the
site phases recur under that rotation. The restrictions of $B_{1}$ and of $E_{1}$ to
$K$ do not contain $\omega$, so that both channels close together. The solutions of
Eq.~(\ref{eq:hexlattice}) are lines parallel to the sixfold axis through a triangular
lattice of spacing $2\pi/(\sqrt{3}R)$ in the plane of the ring, with $R$ the radius of
the ring.

Figure~\ref{fig:hexring} shows two $E_{1}$--$E_{2}$ transitions, computed for the
nearest-neighbor $s=1/2$ Heisenberg ring with $J=1$ (ground-state energy
$E_{0}=-2.8027756$), ring radius $R=3$ and all scattering amplitudes equal to
unity.
The two transitions connect the $E_{1}$ level at $E=-1$ ($S=1$) with the two lowest
$E_{2}$ levels, at $E=-1.2808$ ($S=1$) and at $E=0$ ($S=2$). Both transitions are dark on the extinction lattice of Eq.~(\ref{eq:hexlattice}), and elsewhere they differ. On one quarter of the lines, those with
$\Q\cdot\Rv_{j}\in2\pi\mathbb{Z}$ at every site, every phase factor equals
unity. The amplitude vector is then uniform and lies entirely in the
discarded channel $A_{1}$, so every transition of the ring is dark on these
lines.

\begin{figure*}[t]
\centering
\includegraphics[width=0.72\textwidth]{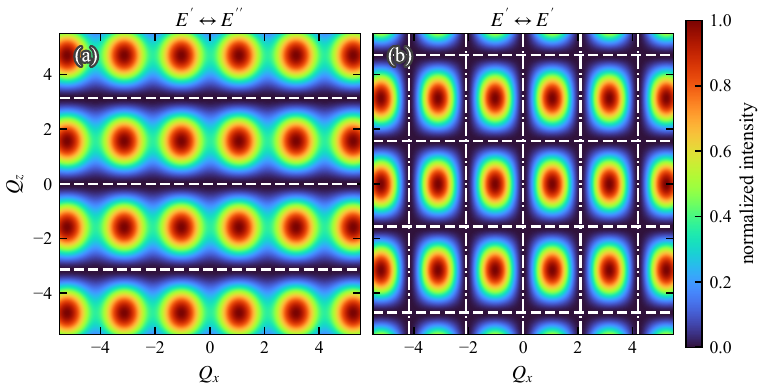}
\caption{\label{fig:prism}Computed intensity of two pairs of levels of the
trigonal prism specified in the text, in the plane $Q_{y}=0$, which contains the threefold axis. Panel (a) is the pair
of the $E'$ level at $E=-1.4297$ ($S=1$) and the $E''$ level at $E=-0.4583$
($S=1$), dark on the planes of Eq.~(\ref{eq:prismplanes}). Panel (b) is the
pair of the two lowest $E'$ levels at $E=-1.4297$ and $E=-0.4869$ (both $S=1$), dark
on the shifted planes of Eq.~(\ref{eq:shiftedplanes}) and on the lattice of
lines of equal triangle phases. Dashed and dash-dotted lines mark these
extinctions.}
\end{figure*}

\subsection{Trigonal prism: extinctions on planes}
\label{sec:prism}

A trigonal prism shows the same recurrence of the site phases
(Sec.~\ref{sec:closedforms}) with a different outcome. Instead of a lattice
of lines, the extinctions form parallel planes. Six equal sites form two
eclipsed equilateral triangles, with $s$ the distance between neighboring
sites of one triangle and $h$ that between each site and its partner in the
other triangle. The point group is $D_{3h}$, and all sites belong to one
orbit. Under the permutations they span
$\Gamma^{(N)}=A_{1}'+E'+A_{2}''+E''$. The operator of $A_{1}'$ is the total spin
$\hat{\mathbf{S}}$, which is discarded, so that the channels are $E'$, $A_{2}''$
and $E''$. One pair does not have a universal function: the product
$E'\otimes E''=A_{1}''+A_{2}''+E''$ contains the site species $A_{2}''$ and $E''$. The two triangles are
eclipsed, so both project onto the same
three positions $\bm{\rho}_{j}$, $j=0,1,2$, with vanishing $z$ component. With the cyclic sums over these positions,
\begin{equation}
\tau_{m}(\Q)=\sum_{j=0}^{2}e^{-2\pi imj/3}\,e^{i\Q\cdot\bm{\rho}_{j}},
\label{eq:prismtau}
\end{equation}
the weights are
\begin{equation}
\Phi_{A_{2}''}^{2}=\tfrac{2}{3}\sin^{2}(\tfrac{1}{2}Q_{z}h)\,\lvert\tau_{0}\rvert^{2},
\label{eq:prismweightA}
\end{equation}
\begin{equation}
\Phi_{E''}^{2}=\tfrac{2}{3}\sin^{2}(\tfrac{1}{2}Q_{z}h)\,
\bigl(\lvert\tau_{1}\rvert^{2}+\lvert\tau_{2}\rvert^{2}\bigr).
\label{eq:prismweightE}
\end{equation}
Both weights contain the factor $\sin^{2}(\tfrac{1}{2}Q_{z}h)$, and the three sums $\tau_{m}$ cannot all vanish, since
$\tau_{0}+\tau_{1}+\tau_{2}=3\,e^{i\Q\cdot\bm{\rho}_{0}}$. The extinction set of
the pair is therefore the family of parallel planes
\begin{equation}
Q_{z}h\in2\pi\mathbb{Z}.
\label{eq:prismplanes}
\end{equation}
This is again Eq.~(\ref{eq:congruence}) with $K=\{E,\sigma_{h}\}$ and $\omega=1$: the
reflection carries each site to its partner in the other triangle, and every
difference vector equals $h\hat{\mathbf{z}}$. Where the difference vectors of the ring of Sec.~\ref{sec:hexring} spanned the plane and gave a lattice of lines, they now span a single direction and give
planes, the case in which a pair with two mediating channels retains a
two-dimensional extinction set, because both weights vanish under one and the same condition.

Figure~\ref{fig:prism}(a) shows one of these transitions, computed for
$s=2\sqrt{3}$, $h=2$ and the Heisenberg Hamiltonian with $J_{s}=1$ within the
triangles, $J_{h}=\tfrac{1}{2}$ on the three vertical edges and
$J_{c}=\tfrac{1}{3}$ on the six remaining bonds between the triangles
(ground-state energy $E_{0}=-1.9721271$). The model without $J_{c}$ is accidentally degenerate at every
$J_{h}$, and the stated value of $J_{c}$ lifts these degeneracies. Every pair of an $E'$ and an $E''$ level shares these extinctions.

In addition, the prism realizes the recurrence of the site phases with a
nontrivial common factor, $\omega=-1$. For the pairs $(E',E')$ and $(E'',E'')$ the only channel is $E'$, since
$E'\otimes E'=E''\otimes E''=A_{1}'+A_{2}'+E'$, $A_{1}'$ is the
discarded total spin and $A_{2}'$ is not a site species. The weight of $E'$
contains the complementary factor of Eq.~(\ref{eq:prismweightE}),
\begin{equation}
\Phi_{E'}^{2}=\tfrac{2}{3}\cos^{2}(\tfrac{1}{2}Q_{z}h)\,
\bigl(\lvert\tau_{1}\rvert^{2}+\lvert\tau_{2}\rvert^{2}\bigr),
\label{eq:eprime}
\end{equation}
and its zero set is the union of two components, one per mechanism of
Eq.~(\ref{eq:congruence}). On the shifted planes
\begin{equation}
Q_{z}h\in\pi+2\pi\mathbb{Z}
\label{eq:shiftedplanes}
\end{equation}
the reflection gives $D(\sigma_{h})\,\mathbf{c}(\Q)=-\mathbf{c}(\Q)$, the case
$K=\{E,\sigma_{h}\}$ with $\omega=-1$: the restriction of $E'$, which is even under
$\sigma_{h}$, does not contain $\omega$, while $A_{2}''$ and $E''$ are odd and stay
open there. Where $\lvert\tau_{1}\rvert^{2}+\lvert\tau_{2}\rvert^{2}$ vanishes instead, the
three phase factors of one triangle are equal,
$e^{i\Q\cdot\bm{\rho}_{0}}=e^{i\Q\cdot\bm{\rho}_{1}}=e^{i\Q\cdot\bm{\rho}_{2}}$.
This happens on lines parallel to the threefold axis through a triangular
lattice of spacing $4\pi/(\sqrt{3}\,s)$ and corresponds to
Eq.~(\ref{eq:congruence}) with $K=C_{3}$ and $\omega=1$. On these lines every
channel except $A_{2}''$ closes. Where the lines meet the planes of
Eq.~(\ref{eq:prismplanes}), $A_{2}''$ closes as well.
Figure~\ref{fig:prism}(b) shows the pair $(E',E')$; the pair $(E'',E'')$ gives the
same normalized map, the case of Eq.~(\ref{eq:factor}), and both pairs are dark on the shifted planes and on the lines. The extinctions
are complementary to those of the pair $(E',E'')$: on the planes of
Eq.~(\ref{eq:prismplanes}) the pairs $(E',E')$ and $(E'',E'')$ stay bright away from the lines of
equal triangle phases, on the shifted planes of
Eq.~(\ref{eq:shiftedplanes}) the pair $(E',E'')$ does, and the two level species determine which family of planes is dark.

\begin{figure*}[t]
\centering
\includegraphics[width=0.689\textwidth]{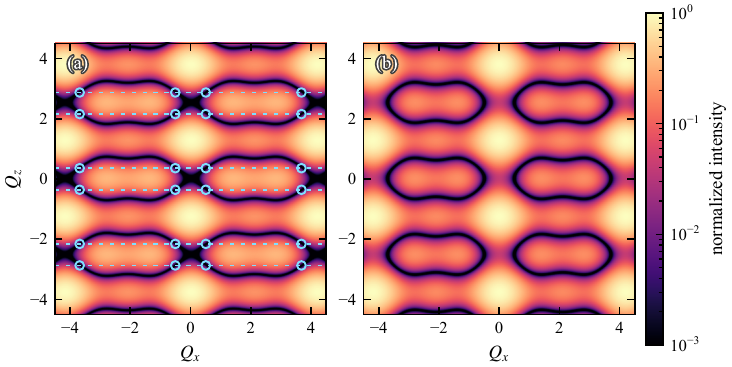}
\caption{\label{fig:zbipy}Computed intensity of the transition between the
two lowest $A_{1g}$ levels of the centered bipyramid specified in the text, in the plane $Q_{y}=0$. In (a) the scattering amplitudes are the constants $(1,0.8,0.5)$ for ring,
apexes and center. Dashed lines mark the planes $\cos(hQ_{z})=F_{c}/F_{a}$,
and circles their crossings with $F_{r}\bar{c}_{r}=F_{a}\cos(hQ_{z})$, the
extinctions of Eq.~(\ref{eq:zbipy}). The other dark curves of panel (a)
are accidental zeros of this Hamiltonian. In (b) the amplitudes are the constants
$(1,0.618,0.8)$, so that $F_{c}>F_{a}$: the condition $\cos(hQ_{z})=F_{c}/F_{a}$ has no solution, no extinction exists, and the dark rings are accidental zeros of this
particular Hamiltonian.}
\end{figure*}

\subsection{Centered hexagon and centered bipyramid: extinctions set by
the amplitudes}
\label{sec:zbipy}

In the clusters treated so far the extinctions followed from the geometry
alone. This subsection adds a third orbit and shows that the extinctions
then come to depend on the scattering amplitudes, in position and in
existence. Two clusters carry the argument. In the centered hexagon the
amplitudes decide, for one kind of transition, whether extinctions exist at
all. In the centered bipyramid they set where the extinctions lie and
whether they exist.

The first cluster is a centered hexagon. Six sites of two alternating kinds
on a circle form two triangle orbits, with scattering amplitudes $F_{1}$
and $F_{2}$, and a central site with amplitude $F_{c}$ is the third orbit.
The point group is $D_{3h}$. The cluster is planar, so the permutations
form $D_{3}$, and the sites span $\Gamma^{(N)}=3A_{1}+2E$. With the total
spin discarded, both channels, $A_{1}$ and $E$, hold two occurrences, and
no pair of levels has a universal function. One pair,
$(A_{1},A_{2})$, admits no channel at all and is dark at every momentum
transfer, since $A_{2}$ is not a site species.

The amplitudes do not reach every pair. The pairs of level species whose
only channel is $E$ are dark on a lattice of lines, by the recurrence of
Sec.~\ref{sec:closedforms}. The lines are those of
Eq.~(\ref{eq:congruence}) with $K=C_{3}$ and $\omega=1$, the kind met at
the trigonal prism of Sec.~\ref{sec:prism}. On them the phases within each
triangle agree, i.e.\ $\Q\cdot(\Rv_{j}-\Rv_{k})\in2\pi\mathbb{Z}$ for
the sites of each triangle, and the amplitude vector is constant on each
triangle. It then lies entirely in the totally symmetric channel, and the
channel $E$ closes. These lines follow from the congruences alone and stay
in place for any choice of the amplitudes.

The pairs that admit the totally symmetric channel are the ones the
amplitudes govern. That channel closes where the three orbit means of
Eq.~(\ref{eq:orbitiso}) share one value. The sites of the second triangle
are opposite to those of the first, so its mean phase factor is the complex
conjugate of that of the first, and the closing condition reads
\begin{equation}
F_{1}C=F_{2}C^{*}=F_{c},
\label{eq:hexmeans}
\end{equation}
with $C=\tfrac{1}{3}\sum_{j\in T_{1}}e^{i\Q\cdot\Rv_{j}}$ the mean phase
factor of the first triangle $T_{1}$. Equation~(\ref{eq:hexmeans}) forces
$C$ to be real and equal to both $F_{c}/F_{1}$ and $F_{c}/F_{2}$, which
fails at every momentum transfer when $F_{1}\neq F_{2}$ and $F_{c}\neq0$.
These pairs therefore have no extinctions at all.

The second cluster, the centered bipyramid, lifts the third orbit out of
the plane and thereby recovers extinctions, in a form that no other
mechanism of this work produces. We take six equal sites on a circle of
radius $R$, one site at $+h$ and one at $-h$ on the sixfold axis, and one
site at the center, with the point group $D_{6h}$. The three orbits are the
ring, the axial pair and the center, with scattering amplitudes $F_{r}$,
$F_{a}$ and $F_{c}$, and each orbit is centrosymmetric. The sites span
$\Gamma^{(N)}=3A_{1g}+A_{2u}+B_{1u}+E_{1u}+E_{2g}$, and the totally
symmetric channel retains two occurrences after the total spin is
discarded. A pair of two levels of the same one-dimensional species admits
only that channel, since the product of such a species with itself is
totally symmetric.

As in the hexagon, the channel closes where the three orbit means of
Eq.~(\ref{eq:orbitiso}) share one value. Every orbit is centrosymmetric, so
each mean is real. The ring contributes $F_{r}\,\bar{c}_{r}(\Q)$, with
$\bar{c}_{r}$ the mean cosine over the ring, the axial pair contributes
$F_{a}\cos(hQ_{z})$, and the center contributes $F_{c}$. Two real
conditions remain,
\begin{equation}
F_{r}\,\bar{c}_{r}(\Q)=F_{a}\cos(hQ_{z})=F_{c},
\label{eq:zbipy}
\end{equation}
and the extinction set is one-dimensional: closed curves on which
$\bar{c}_{r}=F_{c}/F_{r}$, in the planes where $\cos(hQ_{z})=F_{c}/F_{a}$.
Neither condition need have a solution. For $|F_{c}/F_{a}|>1$ the second
condition of Eq.~(\ref{eq:zbipy}) has no solution and the pairs have no
extinction. The first condition fails likewise
when $F_{c}/F_{r}$ lies outside the range $[-\tfrac{1}{2},1]$ of
$\bar{c}_{r}$, and at the two endpoints of that range the curves
collapse to isolated points. If the amplitudes
are $\Q$-dependent form factors instead of constants, the two conditions of
Eq.~(\ref{eq:zbipy}) acquire a dependence on the magnitude of the momentum
transfer. The planes bend, and the extinctions move.

Figure~\ref{fig:zbipy} shows the transition between the two lowest $A_{1g}$
levels, computed for ring radius $3$, apexes at $z=\pm2.5$ and the
Heisenberg model with coupling $1$ on the nearest-neighbor ring bonds,
$1/2$ between ring and apexes, $1/3$ between center and ring and $1/4$
between center and apexes. The ground-state energy is $E_{0}=-3.1335457$,
and the two levels lie at $E=-3.1335$ and $E=-2.8942$, both with $S=1/2$.
With the constant amplitudes $(F_{r},F_{a},F_{c})=(1,0.8,0.5)$ of panel (a)
the branches of $\cos(hQ_{z})=F_{c}/F_{a}$ are straight lines, and the
computed maps of this and of two further pairs of $A_{1g}$ levels are dark
where the plane of the figure cuts the curves. Panel (b) keeps constant
amplitudes but sets them to $(1,0.618,0.8)$, so that $F_{c}>F_{a}$. No
momentum transfer then satisfies Eq.~(\ref{eq:zbipy}), the indicator
vanishes nowhere, and the same pairs have no extinction. The intensity of
one particular Hamiltonian can still vanish accidentally, by cancellation
between the two occurrences of the mediating species. The dark rings of
Fig.~\ref{fig:zbipy}(b) are such accidental zeros, and they move when the
parameters change.

\begin{figure}[t]
\includegraphics[width=\columnwidth]{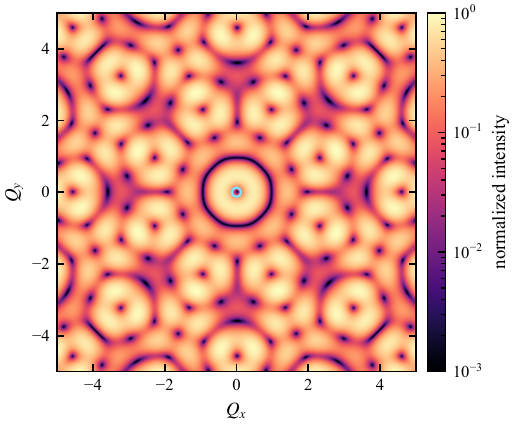}
\caption{\label{fig:ringuni}Computed intensity of the $\Q$-universal
transition between the lowest $A_{1}$ level ($S=0$) and the lowest $E_{1}$
level ($S=1$) of the ring of eight equal sites specified in the text, in the plane $Q_{z}=1.3$. The circle marks the principal axis. The extinction covers the entire $Q_{z}$ axis.}
\end{figure}

\begin{figure*}[t]
\includegraphics[width=\textwidth]{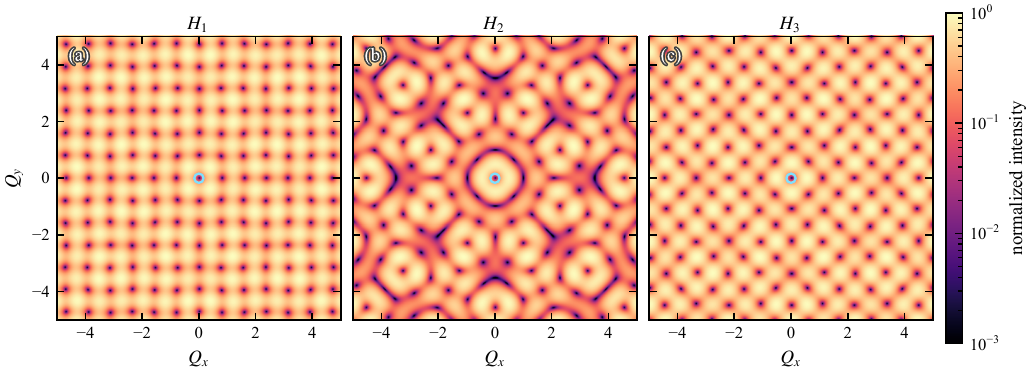}
\caption{\label{fig:ringalt}Computed intensity of the transition between the
lowest $A_{1}$ level ($S=0$) and the lowest $E$ level ($S=1$) of the ring of
two alternating sites, in the plane $Q_{z}=1.3$, for the three invariant
Hamiltonians $H_{1}$, $H_{2}$ and $H_{3}$ stated in the text. The mediating channel has effective
multiplicity two. Only the principal axis (circles) is a systematic
extinction in every map (the dark dots are accidental zeros whose positions
depend on the Hamiltonian parameters).}
\end{figure*}

\begin{figure}[t]
\includegraphics[width=\columnwidth]{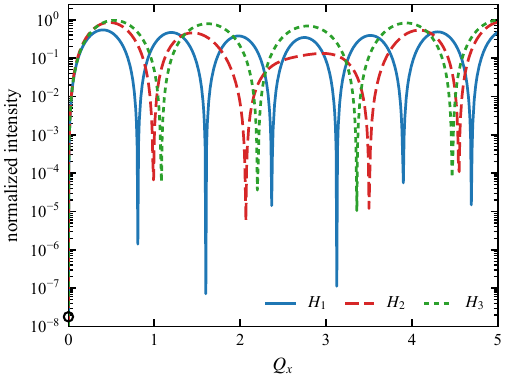}
\caption{\label{fig:ringcut}Intensity of the transition of
Fig.~\ref{fig:ringalt} for the three Hamiltonians, along $Q_{x}$ at $Q_{y}=0$, $Q_{z}=1.3$, each curve
divided by the largest value of its map. The curves are symmetric under
$Q_{x}\to-Q_{x}$, so only $Q_{x}\geq0$ is shown. On the axis (circle at
$Q_{x}=0$) the intensity vanishes for all three Hamiltonians.}
\end{figure}

\subsection{Alternating ring: extinctions versus accidental zeros}
\label{sec:ring}

Rings in which two kinds of magnetic site alternate are common, whether the
two kinds differ in the metal or in the bridging
ligand~\cite{Caciuffo2005,Craig2019,Gatteschi2003,Schnack2019}. We compare
a ring of eight equal sites with a ring of two alternating kinds on the
same circle. The comparison makes the loss of the universal $\Q$ dependence
directly visible, and it forces the distinction that any use of the theory
must draw: which zeros of a map are extinctions and which are accidental.
The anisotropic description then adds extinctions that lie away from every
symmetry element of the cluster.

Both rings have eight sites on a circle of radius $R$. In the first all
sites are equivalent. The point group is $D_{8h}$, and the couplings below
are $D_{8}$-invariant. In the second, two kinds of site alternate, so that
the sites fall into two orbits of four. The point group is $D_{4h}$, and
the scattering amplitude takes the value $F_{1}$ on the first orbit and
$F_{2}$ on the second. All eight sites lie in one plane, so that
$\sigma_{h}$ leaves each of them in place. The permutations of the
isotropic description therefore form $D_{8}$ for the equal ring and $D_{4}$
for the alternating one, and the species of these groups label the levels.
The sites of the alternating ring span
$\Gamma^{(N)}=2A_{1}+B_{1}+B_{2}+2E$. The $x$ axis passes through a site of
the first orbit, and $B_{1}$ is the species that is symmetric under the
twofold rotations about the axes through the sites of that orbit.

Figures~\ref{fig:ringuni}--\ref{fig:ringcut} contrast the two rings on one
transition, computed for radius $R=4$ and, for the alternating ring, the
amplitudes $F_{1}=1$ and $F_{2}=0.618$. Figure~\ref{fig:ringuni} shows the
transition between the lowest $A_{1}$ level ($S=0$) and the lowest $E_{1}$
level ($S=1$) of the equal ring. Its mediating channel $E_{1}$ has
effective multiplicity one, and the intensity is the universal function
known for isotropic spin rings~\cite{Waldmann2003,GT2021a}. Under the
reduction from $D_{8}$ to $D_{4}$ the species $A_{1}$ and $E_{1}$ of these
levels turn into $A_{1}$ and $E$, and the corresponding transition of the
alternating ring, between the lowest $A_{1}$ level ($S=0$) and the lowest
$E$ level ($S=1$), is mediated by the channel $E$ of effective multiplicity
two: no universal function is left.

Three Hamiltonians $H_{1}$, $H_{2}$ and $H_{3}$ are used for each ring,
stated as scalar couplings per orbit of symmetry-equivalent bonds. For the
equal ring the couplings on the nearest-neighbor, next-nearest-neighbor and
diameter bonds are $(1,\tfrac{1}{4},\tfrac{1}{3})$,
$(1,\tfrac{2}{3},\tfrac{1}{3})$ and $(1,\tfrac{3}{4},\tfrac{1}{2})$. For
the alternating ring the nearest-neighbor coupling is $1$, and the
couplings
$(J_{2}^{(1)},J_{2}^{(2)},J_{\mathrm{d}}^{(1)},J_{\mathrm{d}}^{(2)})$ on
the two next-nearest-neighbor orbits and the two diameter orbits are
$(\tfrac{2}{3},\tfrac{1}{4},0,0)$, $(\tfrac{1}{4},0,0,\tfrac{1}{2})$ and
$(0,\tfrac{3}{4},\tfrac{1}{4},0)$, all other couplings being zero.

Every Hamiltonian of the equal ring gives the same map of
Fig.~\ref{fig:ringuni} -- the maps computed for $H_{1}$, $H_{2}$ and
$H_{3}$ coincide to numerical precision -- while the three Hamiltonians of
the alternating ring give the three visibly different maps of
Fig.~\ref{fig:ringalt}. The point does not depend on varying the
Hamiltonian. For a single Hamiltonian, different level pairs of the same
pair of species give different maps as well.

The maps lie in the plane $Q_{z}=1.3$. The $Q_{z}$ axis is an extinction of this transition for both rings -- for the equal ring through the universal function, for
the alternating ring as the closed form below shows -- and every map is
dark where the plane crosses it. Figure~\ref{fig:ringcut} follows the three
maps of Fig.~\ref{fig:ringalt} along the line $Q_{y}=0$. On the axis the
intensity vanishes for all three Hamiltonians. The other minima along the
cut are exact zeros as well, at Hamiltonian-dependent positions --
accidental zeros that a single map cannot tell from extinctions. They are
exact because the cut lies in a mirror plane of the cluster, where the transition amplitude reduces to a single real function. Its
sign changes force the intensity through zero.

So far one transition of the alternating ring was followed. In all, thirteen pairs of level species admit at least one channel, in
five different sets of channels. Three of the sets consist of a single
channel of effective multiplicity one and belong to the treatment of
Refs.~\onlinecite{GT2021a,GT2021b}, with the zeros of the tabulated
functions as their extinctions. The remaining two sets are the channel
$E$ of effective multiplicity two, and the combination of $B_{1}$, $B_{2}$
and $A_{1}$ admitted by the pair $(E,E)$. The weight of the channel $E$
follows from the projection in closed form,
\begin{equation}
\Phi_{E}^{2}=2\bigl[F_{1}^{2}\bigl(\sin^{2}RQ_{x}+\sin^{2}RQ_{y}\bigr)
+F_{2}^{2}\bigl(\sin^{2}u+\sin^{2}v\bigr)\bigr],
\label{eq:eweight}
\end{equation}
with $u=R(Q_{x}+Q_{y})/\sqrt{2}$ and $v=R(Q_{y}-Q_{x})/\sqrt{2}$ the
coordinates turned by an eighth of a turn. This expression and the closed
forms below are sums of squares with one amplitude prefactor per orbit, so
their zeros are unaffected when the orbits differ in their form factors, as
long as no form factor vanishes.

For the pairs whose only channel is $E$, the extinctions are where the four
squares of Eq.~(\ref{eq:eweight}) all vanish, which requires $RQ_{x}$,
$RQ_{y}$ and the two turned coordinates to be multiples of $\pi$. Since
$\sqrt{2}$ is irrational this happens on the fourfold axis alone, and the
surfaces and lattices of the preceding subsections have become a line. The
pair $(E,E)$ faces three conditions -- one per mediating channel -- on the
two in-plane components, and like the corresponding pairs of the centered
hexagon of Sec.~\ref{sec:zbipy} no extinction was found for it.

Whether an extinction set can be seen depends on how steeply the intensity
rises beside it. The intensity of a transition that is dark on a rotation
axis rises as $q^{\mathcal{P}}$ with the distance $q$ from the axis, and
the symmetry species of the two levels give a lower bound for
$\mathcal{P}$ (Supplemental Material~\cite{suppmat}). Beside the fourfold
axis of the alternating ring the pair $(A_{1},E)$ rises as $q^{2}$ and the
pair $(A_{1},B_{1})$ as $q^{4}$, with measured exponents $1.998$ and
$3.999$. The exponent determines how wide the region is in which the
intensity stays below a given fraction of its value away from the axis,
which is the quantity that a measurement of finite resolution registers.
Of two conjectured assignments the one with the larger exponent is the
easier to test.

The anisotropic description multiplies this bookkeeping -- the same ring
then has $47$ pairs of level species in eleven sets of channels -- and it
opens extinctions that detach from every symmetry element. We follow the
pairs whose only channel is $A_{2g}$, the species of the spin component
along the fourfold axis. Each of the two orbits is a square of four sites
and contributes one occurrence, so that Eq.~(\ref{eq:orbitaniso}) applies
and the channel closes where both orbit sums vanish,
\begin{equation}
\cos RQ_{x}+\cos RQ_{y}=0
\quad\text{and}\quad
\cos u+\cos v=0,
\label{eq:octlines}
\end{equation}
with $u$ and $v$ as in Eq.~(\ref{eq:eweight}). Writing $C_{1}$ and $C_{2}$
for the two sums of Eq.~(\ref{eq:octlines}), the orbit sums are
$t_{1}=2F_{1}C_{1}$ and $t_{2}=2F_{2}C_{2}$, and the channel weight of
Eq.~(\ref{eq:orbitaniso}) is
$\mathcal{K}_{A_{2g}}=\sin^{2}\theta_{\Q}\bigl[F_{1}^{2}C_{1}^{2}
+F_{2}^{2}C_{2}^{2}\bigr]$. The factor $\sin^{2}\theta_{\Q}$ vanishes on
the $Q_{z}$ axis, as for every channel of the axial spin component.

Away from the axis the channel closes only where both conditions of
Eq.~(\ref{eq:octlines}) hold at once. Each condition on its own is a
family of straight lines in the plane -- for the first family the
diagonals $Q_{x}\pm Q_{y}=(2n+1)\pi/R$, for the second the lines
$Q_{x}=(2n+1)\pi/(\sqrt{2}R)$ and $Q_{y}=(2n+1)\pi/(\sqrt{2}R)$, with
integer $n$ -- and the extinctions are the crossings of the two families,
in three dimensions a set of lines parallel to the axis. The set of
crossings is not periodic. The two families share the spacing
$\sqrt{2}\,\pi/R$ between neighboring lines but run at $45^{\circ}$ to
each other, so that along a coordinate axis they recur with the
incommensurate periods $2\pi/R$ and $\sqrt{2}\,\pi/R$, whose ratio is the
same irrational number that reduces the $E$ set to the axis.

\begin{figure}[t]
\includegraphics[width=\columnwidth]{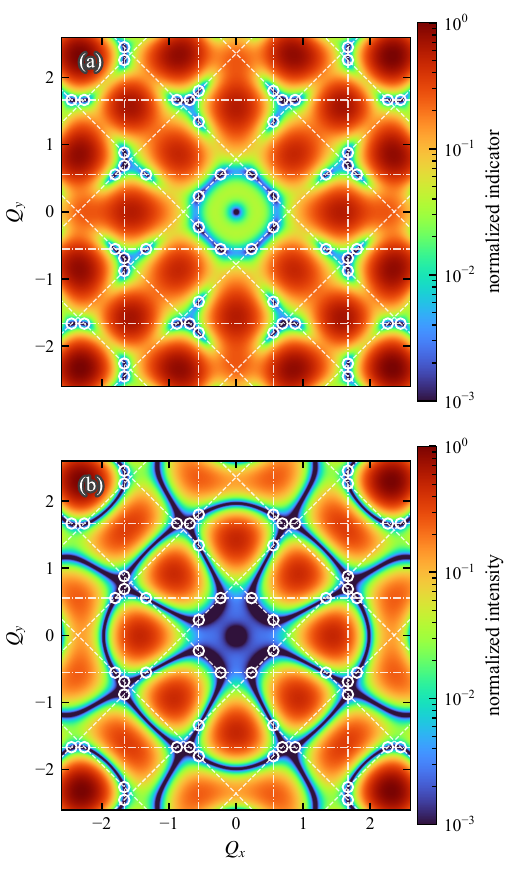}
\caption{\label{fig:a2gring}Channel $A_{2g}$ of the ring of two
alternating sites, anisotropic description, in the plane $Q_{z}=1.1$.
Panel (a) shows the channel weight $\mathcal{K}_{A_{2g}}$, panel (b) the
computed intensity of the transition between the lowest $A_{1g}$ and
$A_{2g}$ levels, for the anisotropic Hamiltonian stated in the text.
Dashed lines mark the zeros of the first sum $C_{1}$ of Eq.~(\ref{eq:octlines}), dash-dotted lines those of the second sum $C_{2}$. Both orbit sums vanish only at the crossings of the two families. Away from the axis, only the crossings, marked by circles, are extinctions. The central dark spot lies on the
$Q_{z}$ axis, where $\sin^{2}\theta_{\Q}$ vanishes. The additional dark
curves in (b) are accidental zeros that move with the Hamiltonian.}
\end{figure}

Figure~\ref{fig:a2gring} compares the channel weight with the computed
intensity of the pair of the lowest $A_{1g}$ and $A_{2g}$ levels in the
plane $Q_{z}=1.1$, for an anisotropic Hamiltonian that keeps the isotropic
nearest-neighbor coupling $1$ and places anisotropic tensors on the two
diameter orbits, in Cartesian coordinates
$\operatorname{diag}(\tfrac{1}{2},\tfrac{1}{4},\tfrac{3}{4})$ on the
diameters through the first orbit, and principal values
$(\tfrac{1}{4},\tfrac{1}{2},\tfrac{1}{4})$ along bond direction, in-plane
normal and $z$ on the others. The crossings have a trivial stabilizer
$H(\Q)$, so these extinctions lie away from every symmetry element of the
cluster. The computed intensity vanishes at every crossing of the plane
shown, and in addition on curves of accidental zeros that pass through the
crossings and move with the Hamiltonian.

A trivial stabilizer does not require two incommensurate families. In the
butterfly arrangement at the core of a large family of tetranuclear
compounds~\cite{McCusker1991,Cauchy2006} -- two orbits of two sites each,
with the point group $C_{2v}$ -- the difference vectors span two
directions, and Eq.~(\ref{eq:congruence}) with $K=G$ and $\omega=1$ alone
gives a lattice of lines parallel to the twofold axis. In the
isotropic description every channel except the totally symmetric one
closes on these lines. Away from the two mirror planes, the lines stand
clear of every symmetry element of the cluster.

\begin{figure*}[t]
\centering
\includegraphics[width=0.565\textwidth]{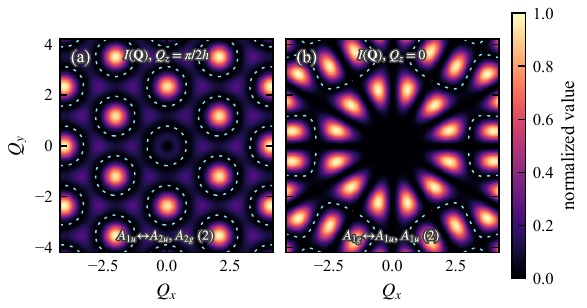}
\caption{\label{fig:bahn}Hexagonal bipyramid, anisotropic description,
for the Hamiltonian stated in the text. Panel (a) shows the computed
intensity of one of the four pairs whose only channel is $A_{2g}$,
between the $A_{1u}$ level at $E=-1.7656$ and the $A_{2u}$ level at
$E=-1.3029$, in the plane $hQ_{z}=\pi/2$. Panel (b) shows the computed
intensity of the pair of the $A_{1g}$ level at $E=-2.6418$ and the
$A_{1u}$ level at $E=-1.7656$, whose only channel is $A_{1u}$, in the
plane $Q_{z}=0$. Dashed curves mark the predicted extinctions, which
follow from the site positions without any Hamiltonian input: the zeros
of the ring sum in (a) and, in (b), the curved arcs among the zeros of
Eq.~(\ref{eq:master}). The straight rays of the same zero set appear as
dark spokes and are left unmarked.}
\end{figure*}

\subsection{Hexagonal bipyramid: zeros beyond the closed forms}
\label{sec:bipy}

A hexagonal bipyramid, a ring with one further magnetic site above and
one below its plane, is the richest cluster of this work, and it has
extinctions that no closed form of Sec.~\ref{sec:closedforms} gives. We
take six sites on a circle of radius $R$ together with one site at $+h$
and one at $-h$ on the sixfold axis, with the point group $D_{6h}$. The
principal axis again lies along $z$ and the $x$ axis passes through a
ring site, so that the levels are labeled by the convention of
Sec.~\ref{sec:ring}. Under the permutations the eight sites span
$\Gamma^{(N)}=2A_{1g}+A_{2u}+B_{1u}+E_{1u}+E_{2g}$, and the three spin
components contribute $\Gamma_{\mathrm{ax}}=A_{2g}+E_{1g}$.

In the anisotropic description $66$ pairs of level species admit at
least one channel, in fourteen different sets of channels, and the
isotropic description of the same cluster produces seven sets of its
own. Together these twenty-one sets of channels yield a nonempty
extinction set in nineteen cases, sixteen of them distinct. The two
sets without any extinction are the channel $E_{1g}$ alone, whose
weight never vanishes, and the larger set that contains it. Two
channels are followed here. The first, $A_{2g}$, shows how the orbit
sums place curves inside special planes. The second, $A_{1u}$, has the
zeros that no closed form gives, found only by the full indicator.

The pairs whose only channel is $A_{2g}$, the species of the spin
component along the axis, are $(A_{1u},A_{2u})$, $(B_{2g},B_{1g})$,
$(B_{1u},B_{2u})$ and $(A_{2g},A_{1g})$. Each of the two orbits
contributes one occurrence of $A_{2g}$, so that
Eq.~(\ref{eq:orbitaniso}) applies and the channel closes where both
orbit sums vanish separately -- unlike at the centered bipyramid of
Sec.~\ref{sec:zbipy}, where the orbit means had to balance against each
other. The axial sum is $t_{a}=2F_{a}\cos(hQ_{z})$, with $F_{a}$ the
amplitude of the two apexes. It vanishes on the equally spaced planes
\begin{equation}
hQ_{z}\in\tfrac{\pi}{2}+\pi\mathbb{Z},
\label{eq:bipyplanes}
\end{equation}
and the ring sum $t_{r}$ of the six ring sites, the $k=0$ sum of
Eq.~(\ref{eq:hexsums}) with the ring's own amplitude, has
closed curves as its zeros. The six ring positions are the
nearest-neighbor vectors of a triangular lattice of constant $R$, so
the ring sum is periodic in the in-plane component of the momentum
transfer, with magnitude $q$ and azimuth $\varphi_{q}$, and its zero
curves are congruent copies of the curve around the origin, repeated on
the reciprocal triangular lattice. Each copy is nearly but not exactly
a circle: the Jacobi--Anger expansion of the Supplemental
Material~\cite{suppmat} turns the ring sum, up to the factor $6F_{r}$,
into the Bessel-function series
$J_{0}(qR)+2\sum_{k\geq1}(-1)^{k}J_{6k}(qR)\cos(6k\varphi_{q})$, so
each curve follows the first zero of $J_{0}$, with a sixfold modulation
of about half a percent of its radius. The sum of the two orbit contributions is
$\mathcal{K}_{A_{2g}}=\sin^{2}\theta_{\Q}\bigl[6\lvert\bar{t}_{r}\rvert^{2}
+2\lvert\bar{t}_{a}\rvert^{2}\bigr]$, with $\bar{t}_{r}$ and
$\bar{t}_{a}$ the orbit means of the ring and of the axial pair.

Within one plane of Eq.~(\ref{eq:bipyplanes}) the axial condition holds
throughout, so that the ring condition alone remains and the extinction
is a family of closed curves rather than the isolated points in which a plane in general
position meets the extinction set. Figure~\ref{fig:bahn}(a) shows the computed intensity of one of the four transitions in the plane $hQ_{z}=\pi/2$, for ring radius $3.1$ and
apexes at $z=\pm2.5$. The Hamiltonian is invariant and has nonzero
tensors on two bond orbits. On a nearest-neighbor ring bond, in the
frame of radial direction, bond direction and $z$, the tensor has the
symmetric part $\operatorname{diag}(1,\tfrac{3}{4},\tfrac{1}{2})$ and
the Dzyaloshinskii--Moriya $z$ component $\tfrac{1}{4}$. On a
ring--apex bond, in Cartesian coordinates, the tensor is the scalar
$\tfrac{1}{2}$ plus a symmetric $xz$ entry $\tfrac{1}{4}$. All other
couplings are zero, and the remaining bonds are generated by symmetry.

In this plane the axial sum vanishes identically, and the indicator is
then analytically identical, up to a constant factor, to the intensity
of each of the four pairs. It is therefore not plotted separately. At
$hQ_{z}=\pi/4$, where the axial sum is
$\sqrt{2}\,F_{a}$ instead of zero, the same curve stays bright, because the extinction of panel (a) requires both orbit sums. There the four maps no
longer coincide -- they differ in overall strength and, after
normalization, in shape -- and the dark curves there move with the
Hamiltonian while the curve of panel (a) does not.

In the plane of panel (a) the ring condition was solved along
$37$ azimuths spanning $60^{\circ}$, which gives $106$ momentum
transfers on the curve within the window of Fig.~\ref{fig:bahn}(a). There
the computed intensity of each of the four transitions vanishes, and
$98$ of the $106$ momentum transfers have a trivial stabilizer $H(\Q)$
and stand more than a degree from every rotation axis and mirror plane,
so that the extinction is away from every symmetry element of the
cluster.

The pairs whose only channel is $A_{1u}$ are again four, namely
$(B_{2g},B_{2u})$, $(B_{1g},B_{1u})$, $(A_{1g},A_{1u})$ and
$(A_{2g},A_{2u})$. The two occurrences of $A_{1u}$ belong to different
orbits, one to the axial pair with the spin component along the axis
and one to the ring with the spin components in the plane. The first
occurrence has the factor $\sin(hQ_{z})$, from the opposite phases
$e^{\pm ihQ_{z}}$ of the two apexes, and vanishes throughout the plane
$Q_{z}=0$, so that within that plane the ring occurrence alone remains.
Its zero set there consists of twelve straight rays, where the six
vertical mirror planes intersect the plane, together with curved arcs,
one in each of the twelve sectors between neighboring rays and
recurring at larger magnitudes of the momentum transfer.

The two parts of the zero set follow by different routes. The twelve innermost arcs are shown in Fig.~\ref{fig:bahn}(b) and were obtained by
tracing the zeros of Eq.~(\ref{eq:master}), which depend only on the
site positions. For the rays the shorter route is the character count
of Eq.~(\ref{eq:nu}). A momentum transfer in a vertical mirror plane
with $Q_{z}=0$ has a stabilizer of order four, the product of the two
level species restricted to that stabilizer does not contain
$\Gamma_{\perp}$, and $\nu_{nm}$ therefore vanishes.

The arcs lie beyond the reach of the character condition. The
stabilizer has order two on the arcs and at the brightest momentum
transfers of the same plane alike, so Eq.~(\ref{eq:nu}) does not
distinguish dark from bright there. The arcs are genuinely curved.

On the rays and on the arcs the computed intensity of each of the
four transitions vanishes, for the Hamiltonian of Fig.~\ref{fig:bahn}
and for three further invariant sets that place anisotropic tensors on
the ring--apex bonds, on the ring diameters or on the ring next-nearest
bonds. The arcs are thus dark for all four Hamiltonians,
at momentum transfers that no closed form of Sec.~\ref{sec:closedforms}
and no character count predicts -- the case that
Eq.~(\ref{eq:master}) alone settles.

\section{Conclusions}

When several symmetry species mediate a transition of a molecular spin cluster, or
one species occurs more than once among the local spin operators, the point group no
longer fixes the full $\Q$ dependence of the intensity. What it still fixes are momentum transfers at which the intensity vanishes for every Hamiltonian that the
point group admits.

Equations~(\ref{eq:masteriso}) and (\ref{eq:master}) give that part of them which follows from the point group alone, while a transition can be dark at further momentum transfers that they do not account for. The symmetry species of the two
levels enter through a character sum and the geometry through a sum of phase factors over
the site permutations, and the two appear as separate factors in every term of that sum. When each
orbit of magnetic sites contributes exactly one occurrence of the mediating species, the projection separates into
one term per orbit, so that the channel of the spin component along the principal axis is
dark where every orbit sum vanishes and the totally symmetric channel of the isotropic
description is dark where all orbit means take the same value. Where the site
phases recur under a subgroup up to a common factor, the condition becomes a set of linear
congruences, and the extinctions form families of parallel planes or lines. Rotation axes
and mirror planes are one instance of a unit factor, and a lattice of lines at which the
momentum transfer has no symmetry of its own is another.

The extinctions obtained in this way lie on families of parallel planes, on curves and on
lattices of lines, and where three conditions meet they are isolated points.
Across the clusters treated here, a curved surface occurred only where a single species mediates the transition and the local spin operators contain one occurrence
of that species. An experiment must
therefore resolve the direction of the momentum transfer as well as its magnitude. Since the position of an
extinction is fixed by the symmetry species of the two levels, an observed extinction
constrains the species of an excited level once the
species of the initial level is known. The extinctions carry the separation of a dynamic from a geometric factor
into the regime where that separation fails for the intensity itself: the
geometry no longer fixes the whole $\Q$ dependence, but it still fixes where the intensity vanishes.

Whether an extinction can be observed depends on one further condition, which
lies outside the point group of a single molecule. A single-crystal measurement
adds the intensities of all molecules in the crystal, and the sum is dark only
where every molecule is dark. Molecules in different orientations have their
extinctions at different momentum transfers, so the crystal must contain them in one orientation. Orientations related by inversion are the exception. Where a
crystal does contain several orientations, the statement survives for each of them
separately: at a momentum transfer that is an extinction for one orientation,
that orientation contributes nothing, and the measured intensity comes from the
others alone. An intensity that persists where an extinction is
predicted shows that one of the assumptions behind the
prediction fails -- the point group of
the couplings, the species assigned to the two levels, or the scattering amplitudes
of the sites, including the assumed spherical site densities and scalar $g$ values
-- once the resolution of the instrument, the background, and any superposition of
molecular orientations have been accounted for.

\begin{acknowledgments}
S.G.T. was supported by the Deutsche Forschungsgemeinschaft (DFG) under
Project 535298924.
\end{acknowledgments}

\section*{Use of artificial-intelligence tools}

Generative artificial-intelligence tools were used substantively and are disclosed here. Claude Opus (Anthropic, versions 4.6 to 4.8) and
ChatGPT (OpenAI, GPT-5.5) assisted in developing the initial ideas and derivations.
Manuscript editing, proof development, and the development of the
verification and figure programs were subsequently assisted mainly by Claude Opus 5 and Claude Fable 5 (Anthropic). The algebraic and
numerical claims were additionally checked with ChatGPT (OpenAI,
GPT-5.6 ``Sol'') and Kimi K3 (Moonshot AI), through independent
attempts to reproduce or refute them. The author directed these uses, independently
checked the resulting equations and numerical calculations, rejected unsupported
output, and accepts full
responsibility for the manuscript.


\begin{thebibliography}{99}

\bibitem{Gudel1977} H. U. G\"udel and A. Furrer, Mol. Phys. \textbf{33}, 1335 (1977).

\bibitem{Furrer1977} A. Furrer and H. U. G\"udel, Phys. Rev. Lett. \textbf{39}, 657 (1977).

\bibitem{Furrer2013} A. Furrer and O. Waldmann, Rev. Mod. Phys. \textbf{85}, 367 (2013).

\bibitem{Baker2012} M. L. Baker, T. Guidi, S. Carretta, J. Ollivier, H. Mutka,
H. U. G\"udel, G. A. Timco, E. J. L. McInnes, G. Amoretti, R. E. P. Winpenny, and
P. Santini, Nat. Phys. \textbf{8}, 906 (2012).

\bibitem{Garlatti2017} E. Garlatti, T. Guidi, S. Ansbro, P. Santini, G. Amoretti,
J. Ollivier, H. Mutka, G. Timco, I. J. Vitorica-Yrezabal, G. F. S. Whitehead,
R. E. P. Winpenny, and S. Carretta, Nat. Commun. \textbf{8}, 14543 (2017).

\bibitem{Chiesa2017} A. Chiesa, T. Guidi, S. Carretta, S. Ansbro, G. A. Timco,
I. Vitorica-Yrezabal, E. Garlatti, G. Amoretti, R. E. P. Winpenny, and P. Santini,
Phys. Rev. Lett. \textbf{119}, 217202 (2017).

\bibitem{Garlatti2020} E. Garlatti, L. Tesi, A. Lunghi, M. Atzori, D. J. Voneshen,
P. Santini, S. Sanvito, T. Guidi, R. Sessoli, and S. Carretta,
Nat. Commun. \textbf{11}, 1751 (2020).

\bibitem{Garlatti2019} E. Garlatti, A. Chiesa, T. Guidi, G. Amoretti, P. Santini, and
S. Carretta, Eur. J. Inorg. Chem. \textbf{2019}, 1106 (2019).

\bibitem{Furrer1979} A. Furrer and H. U. G\"udel, J. Magn. Magn. Mater.
\textbf{14}, 256 (1979).

\bibitem{Guedel1979} H. U. G\"udel, A. Stebler, and A. Furrer, Inorg. Chem.
\textbf{18}, 1021 (1979).

\bibitem{Furrer1989} A. Furrer, H. U. G\"udel, H. Blank, and A. Heidemann,
Phys. Rev. Lett. \textbf{62}, 210 (1989).

\bibitem{Haraldsen2005} J. T. Haraldsen, T. Barnes, and J. L. Musfeldt,
Phys. Rev. B \textbf{71}, 064403 (2005).

\bibitem{Haraldsen2016} J. T. Haraldsen, Phys. Rev. B \textbf{94}, 054436 (2016).

\bibitem{Waldmann2003} O. Waldmann, Phys. Rev. B \textbf{68}, 174406 (2003).

\bibitem{Waldmann2007} O. Waldmann, R. Bircher, G. Carver, A. Sieber, H. U. G\"udel, and
H. Mutka, Phys. Rev. B \textbf{75}, 174438 (2007).

\bibitem{GT2021a} S. Ghassemi Tabrizi, Phys. Rev. B \textbf{103}, 214422 (2021).

\bibitem{GT2021b} S. Ghassemi Tabrizi, Phys. Rev. B \textbf{104}, 014416 (2021).

\bibitem{Marshall1971} W. Marshall and S. W. Lovesey, \emph{Theory of Thermal Neutron
Scattering} (Clarendon Press, Oxford, 1971).

\bibitem{Waldmann2000} O. Waldmann, Phys. Rev. B \textbf{61}, 6138 (2000).

\bibitem{Klemm2008} R. A. Klemm and D. V. Efremov, Phys. Rev. B \textbf{77}, 184410 (2008).

\bibitem{Altmann1994} S. L. Altmann and P. Herzig, \emph{Point-Group Theory Tables}
(Clarendon Press, Oxford, 1994).

\bibitem{suppmat} See Supplemental Material for an alternating cube
whose extinctions are isolated points and for the derivation of the power
law that governs the rise of the intensity beside an extinction on a
rotation axis.

\bibitem{Caciuffo2005} R. Caciuffo, T. Guidi, G. Amoretti, S. Carretta,
E. Liviotti, P. Santini, C. Mondelli, G. Timco, C. A. Muryn, and
R. E. P. Winpenny, Phys. Rev. B \textbf{71}, 174407 (2005).

\bibitem{Craig2019} G. A. Craig, G. Velmurugan, C. Wilson, R. Valiente,
G. Rajaraman, and M. Murrie, Inorg. Chem. \textbf{58}, 13815 (2019).

\bibitem{Gatteschi2003} D. Gatteschi and R. Sessoli, Angew. Chem. Int. Ed. \textbf{42},
268 (2003).

\bibitem{Schnack2019} J. Schnack, Contemp. Phys. \textbf{60}, 127 (2019).

\bibitem{McCusker1991} J. K. McCusker, J. B. Vincent, E. A. Schmitt,
M. L. Mino, K. Shin, D. K. Coggin, P. M. Hagen, J. C. Huffman, G. Christou,
and D. N. Hendrickson, J. Am. Chem. Soc. \textbf{113}, 3012 (1991).

\bibitem{Cauchy2006} T. Cauchy, E. Ruiz, and S. Alvarez, J. Am. Chem. Soc.
\textbf{128}, 15722 (2006).

\end{thebibliography}
\end{document}